\documentclass[aps,pre,preprint,showpacs,showkeys,groupedaddress]{revtex4}
\newcommand{\W}{10cm}
\usepackage{graphicx}
\usepackage{amsmath}
\begin{document}
\title{Transport in rough self-affine fractures}
\author{German Drazer}
\email{drazer@mailaps.org}
\author{Joel Koplik}
\email{koplik@sci.ccny.cuny.edu}
\affiliation{Benjamin Levich Institute and Department of Physics,
City College of the City University of New York, New York, NY 10031}
\date{\today}
\begin{abstract}
Transport properties of three-dimensional self-affine rough fractures
are studied by means of an effective-medium analysis and numerical
simulations using the Lattice-Boltzmann method. 
The numerical results show that the effective-medium approximation predicts the
right scaling behavior of the permeability and of the velocity fluctuations,
in terms of the aperture of the fracture, the roughness exponent
and the characteristic length of the fracture surfaces, in the limit of small
separation between surfaces. The permeability of the
fractures is also investigated as a function of the normal and lateral
relative displacements between surfaces, and is shown that it can
be bounded by the permeability of two-dimensional fractures. 
The development of channel-like structures in the velocity field is also numerically 
investigated for different relative displacements between surfaces. Finally, the 
dispersion of tracer particles in the velocity field of the fractures is
investigated by analytic and numerical methods. 
The asymptotic dominant role of the geometric dispersion, due to velocity
fluctuations and their spatial correlations, is shown in
the limit of very small separation between fracture surfaces. 

\end{abstract}
\maketitle
\section{Introduction}
\label{intro}

The transport of fluids in geological media plays a dominant role
in different applications such as subsurface hydrology, hydrocarbon 
recovery and waste storage, and the transport properties of such
media involves a combination of fluid flow at different 
length scales. First, there is transport at the microscopic level
through the pore space of the rock itself. Then, at macroscopic length scales, 
the flow through fractures is, in most cases, the dominant transport mechanism.
Finally, at even larger length scales, the dominant convective transport involves 
the flow through fracture networks. The first case has received considerable attention and
is relatively well understood and, in fact, it can be placed within
the framework of transport in porous media \cite{bear,dullien,sahimi95}.
On the other hand, understanding the flow through single fractures is 
clearly a prerequisite to the investigation or modeling of more
complex cases such as the flow through macroscopic fracture networks. 
Another key transport process present in geological systems,
the transport of tracer particles carried by the fluid also requires 
an adequate understanding of the flow properties, in particular, the statistical 
properties of the velocity field.

Fractures have often been modeled as simple Hele-Shaw cells,  with a cubic relation between 
the volumetric flow rate and the average aperture of the
fracture, usually focusing most of the effort on investigating 
the flow through a macroscopic network of such 
fractures. However, although a typical fractured rock surface appears
fairly smooth, aside from some random roughness, suggesting that 
Poiseuille flow in a straight channel is the appropriate model of flow, laboratory
experiments \cite{dijk99,auradou01} and numerical simulations \cite{brown87,GK1} 
indicate that this classical view of a rock fracture as a straight channel is not 
adequate to describe the fluid flow even at low Reynolds numbers.
In fact, more careful analysis showed, in recent years, that geological fractures
present highly spatially correlated, self-affine surfaces, with a roughness
exponent $\zeta \approx 0.8$ surprisingly constant for different types of rocks 
whether naturally or artificially fractured
\cite{bouchaud90,plouraboue95,plouraboue96,meheust00,auradou01}.
In view of these results, theoretical and numerical studies introduced the
complex geometry of the fractures in order to calculate the 
transport properties of the system. However, the majority of these studies
assumed that the Reynolds (lubrication) approximation is valid so that
the velocity field is given by a Poiseuille flow everywhere, with a 
parabolic velocity profile across the aperture and in the direction of 
the mean flow \cite{plouraboue98,roux98}. 

In a previous work \cite{GK1}, we showed 
that, for two-dimensional narrow-fractures, in which the roughness amplitude is large 
compared to the aperture, the Reynolds approximation is no longer valid, 
and proposed a different approach in which the fracture is divided into 
approximately-straight segments with varying orientation angles. This
approach allowed us to obtain the correction to the flow rate due to 
surface roughness as a function of the aperture and was validated by our 
numerical simulations. (A similar approach was used by 
Oron and Berkowitz to analyze the case of fractures with contacts between
surfaces \cite{oron98}.) In this work, we shall further investigate the case of narrow 
fractures, extending the results to three-dimensional fractures. We shall see
that it is possible to obtain analytic expressions for the permeability
of the fracture and for the velocity fluctuations, in the limit of narrow fractures,
by means of the effective-medium approximation. We shall also investigate
the dispersion of tracer particles advected by the flow field 
within the fracture and its dependence on the aperture.

We shall first, in Sec.~\ref{self-affine}, briefly discuss some basic definitions 
associated with self-affine surfaces and present our model of fractures as the gap between two
perfectly matching self-affine surfaces. The Lattice-Boltzmann method, 
used to numerically simulate the fluid flow through the fractures is briefly
presented in Sec.~\ref{LBM}. Then, in Sec.~\ref{permeability} we shall investigate the 
permeability by means of the effective-medium approximation and
compare the results with our numerical results. Finally, in Sec.~\ref{dispersion}
we investigate the dispersion of tracer particles convected by the flow field
within the fracture and its dependence on the aperture of the fracture.

\section{Self-affine fracture surfaces}
\label{self-affine}

Following the work of Mandelbrot \cite{mandelbrot},
the application of the fractal model to describe surface topography
has become widespread as it has been shown in several
experiments that fracture surfaces, with various
materials and fracturing methods, exhibit statistically 
self-affine scaling properties. 
Particularly relevant to our work are the experiments performed with
naturally fractured rocks \cite{brown85,poon92,schmittbuhl93,power97}.
This self-affine description of fracture surfaces
substantially improves the characterization 
of rough surfaces in terms of surface parameters such as  
the root-mean-square (rms) roughness, rms slope, and
density of peaks, in that it provides the scaling behavior of
these parameters, which are not intrinsic properties of the surface 
but strongly depend on the sample size \cite{brown85,poon92}.

\subsection{Scaling properties of self-affine surfaces}
\label{scale}

We shall briefly review here the mathematical characterization of self-affine surfaces,
and its application to the case of fractured rocks. We consider a rock surface 
without overhangs, as it has been shown that, for natural rock surfaces, overhangs 
are not prevalent \cite{brown85}, whose height is given by a by a single-valued function 
$z(x,y)$, with the coordinates $x$ and $y$ lying in the mean plane of the fracture.

Self-affine surfaces
display scale invariance with different dilation ratios along different
spatial directions, remaining unchanged under the rescaling \cite{feder}
\begin{equation}
x \rightarrow \lambda_1 x \qquad
y \rightarrow \lambda_2 y \qquad
z \rightarrow \lambda_3 z
\end{equation}
Here we consider disordered media, so these scaling laws apply only in an
ensemble or spatial average sense.  Experiment indicates that for many
materials isotropy can be assumed in the mean plane, implying that there
is only one non-trivial exponent relating the dilation ratio
in the mean plane to the scaling in the perpendicular direction, {\em i.e.},
$\lambda_1=\lambda_2\equiv\lambda$, and
$\lambda_3=\lambda^\zeta$, where $\zeta$ is usually referred to as the roughness or 
Hurst exponent \cite{mandelbrot}. Therefore, the surface height is a homogeneous function, 
of degree $\zeta$, on the mean plane coordinates,
\begin{equation}
\label{scaling}
z(x,y) = \lambda^{-\zeta} z(\lambda x, \lambda y).
\end{equation}

A surface with Hurst exponent close to 1 has high 
spatial correlation and, as the Hurst exponent 
decreases, heights of nearby points become more uncorrelated,
becoming totally independent for $\zeta$=0.5.
In most studies the roughness exponent is found to be close
to $\zeta=0.8$, leading to the suggestion that it could be a 
universal value, independent of the material and of the 
fracture mode \cite{bouchaud90}. The experimental values
obtained with several different types of rocks, 
whether naturally or artificially fractured, are consistent with 
this suggestion, with $\zeta$ lying around $\zeta=0.8\pm0.05$ 
\cite{poon92,plouraboue95,plouraboue96,meheust00,auradou01}
(Other values may arise from intergranular 
effects as in the case of sandstone rocks \cite{boffa98}.) 
Therefore, as we shall investigate
transport of fluids in fractured rocks, we will use
a roughness exponent $\zeta=0.8$. Furthermore,
it is worth noting that the roughness exponent of a granitic fault
surface at field length scales of several meters
was found to be $\zeta=0.84$ \cite{schmittbuhl93},
thereby supporting the relevance of experimental results obtained
at laboratory scales. 

\subsection{Characteristic length}
\label{topo}

In contrast with the case of self-similar surfaces,
where the fractal dimension fully characterize them, 
the Hurst exponent is not enough to describe a 
self-affine surface, and, in addition, the amplitude
of the fluctuations in the height of the surface
is to be specified. This amplitude of the fluctuations
is usually expressed in terms of the characteristic length $\ell$,
which is the horizontal distance over which fluctuations in height
have a rms slope of one.  Using the
self-affine scaling law for the vertical fluctuations in surface height,
\begin{equation}
\sigma_z^2(r) =
\left\langle [z(\bar{r}')-z(\bar{r}'+ \bar{r})]^2 \right\rangle \propto
r^{2\zeta}
\end{equation}
and the previously mentioned definition of the characteristic 
length $\ell$, 
\begin{equation}
\sigma_z^2(\ell) = \ell^2,
\end{equation}
we can then obtain the variance of the fluctuations in surface height, over any 
length scale $r$, in terms of the roughness exponent $\zeta$ and the characteristic 
length $\ell$,
\begin{equation}
\label{variance}
\sigma_z^2(r) =
\left\langle [z(\bar{r}')-z(\bar{r}'+ \bar{r})]^2 \right\rangle = 
\ell^2 \left( \frac{r}{\ell} \right)^{2\zeta}.
\end{equation}

In principle, the characteristic length $\ell$ can take any value,
however, for fractured rocks it is usually found to be very small, 
and to lie either below the accessible range of length scales in 
the experiment ($\ell \leq 0.5 \mu$m \cite{poon92}) or
below the self-affine range of the surface \cite{auradou01,boffa00}; 
Therefore, we shall not consider cases in which $\ell$ is within the simulated 
range of length scales.

\subsection{Generation of fracture surfaces}
\label{numericalsurfaces}

The self-affine surfaces are generated by a two-dimensional 
generalization of the Fourier synthesis method previously used by us
for the generation of self-affine curves \cite{GK1}. Initially, 
an $L \times L$ matrix, where each element will represent the height of
the discrete version of the fracture surface, is generated with 
statistically independent, Gaussian distributed, random numbers.
Then, the Fourier transform  of this initial random matrix
is modulated by a power-law high-wave-numbers filter that introduces height-to-height 
correlations \cite{voss85}. That is, if $\tilde{Z}(k)$ is the Fourier transform  
of the initial random matrix, then the Fourier transform of the 
matrix representing the fracture surface heights is chosen to be, 
\begin{equation}
\label{spectrum}
Z(k) = k^{-\zeta-1} \tilde{Z}(k).
\end{equation}
Finally, the whole surface is rescaled in order to get the desired 
characteristic length $\ell$. In this work we shall use $\zeta=0.8$
and $\ell \simeq 3\times10^{-3}$. This numerical method generates 
homogeneous and isotropic surfaces and is preferable instead of 
other numerical methods such as the random addition algorithm
(for a discussion of this point see Ref. \cite{kondev00} section V.A). 

\begin{figure}[htb]
\includegraphics[angle=-90,width=\W]{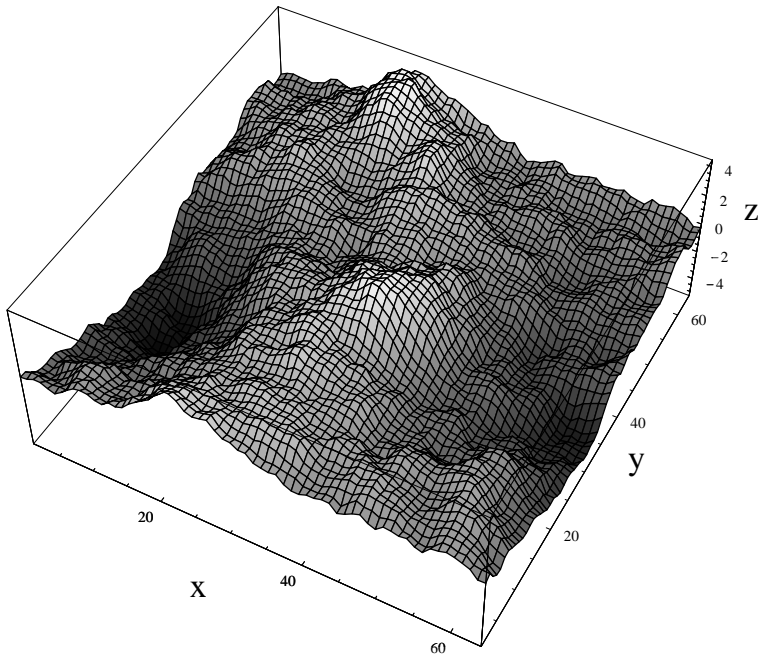}
\caption[surface]{\label{surface} Example of a numerically generated self-affine
surface, with roughness exponent $\zeta=0.8$, characteristic length 
$\ell\simeq3\times10^{-3}$, and size $L\times L= 64 \times 64$.}
\end{figure}

In Fig.~\ref{surface} we show one of the numerically generated
fracture surfaces. The surfaces generated with this method are periodic 
in $x$ and $y$ and, although the periodicity is not physically important 
it has some computational advantages when calculating the
flow field using periodic boundary conditions.

\subsection{Model of fractures}
\label{fractures}

We shall model fractures as the gap between a numerically generated self-affine
surface and its replica, which is translated by a fixed distance $h$ in the direction
normal to the mean plane of the surface. However, during the fracturing process 
of the rock, the opposite matching surfaces might experience a lateral
shift, in addition to the vertical displacement. Therefore, we shall also
consider this situation by introducing a lateral shift to the upper surface
either along the direction of the mean flow or perpendicular to it.

In this work, we will analyze, for both types of fractures, with and without
lateral shift, the case of {\it narrow fractures}, in which the gap size
$h$ is small compared to the vertical fluctuations of the fracture surface
over length scales comparable to the size of the system $L$. In view of
Eq.~(\ref{variance}) we obtain that fluctuations of the surface
height over the whole system are given by 
$\sigma_z(L) = \ell (L/\ell)^{\zeta}$ and, then the limiting
situation of narrow fractures corresponds to,
\begin{equation}
\label{narrow}
\left( \frac{h}{\ell} \right) \ll \left( \frac{L}{\ell} \right)^{\zeta}.
\end{equation}

In our simulations, and for systems size $L=512$ the largest gap
size that fulfills the previous relation is $h_{max} \sim 50$.

\section{Lattice-Boltzmann method}
\label{LBM}

The Lattice-Boltzmann Method (LBM) \cite{rothman,wolf} is particularly 
suitable to investigate the flow of fluids in a highly irregular geometries,
and, specifically, to study the various features of transport
in self-affine rough fractures which are sensitive to their
complex geometrical structure.
In this algorithm, fictitious particles move between
neighboring sites on a lattice, with suitable collision rules, and the 
boundary of the flow domain is simply a surface of sites where the rule 
is modified in some way to keep the particles out.  We use the 
Face Centered Hypercubic (FCHC-projected) version of the LBM,
with a cubic lattice in 3 dimensions and 19 velocities
(D3Q19 in the terminology of \cite{qian}). 
The collision operator is  approximated by a single relaxation time, 
the Bhatnagar-Gross-Krook model \cite{BGK}, and the local equilibrium 
distribution given in \cite{qian} is used. This pseudo-equilibrium distribution locally
preserves mass and momentum values, and is formulated specifically to 
recover the Navier-Stokes equation at large length and time scales.
For the no-slip solid boundaries, we use the simplest implementation of 
particle exclusion -- the bounce-back rule, in which a particle incident on
the boundary reverses its direction.  
Periodic boundary conditions are used for 
the inflow and outflow surfaces.  A constant pressure gradient forcing the 
fluid is added in the $x$ direction, while the gap between surfaces extends 
over $z$.

In what follows all quantities will be rendered dimensionless
by the characteristic units of the numerical simulation.
That is, taking the lattice-spacing as the unit length
and the simulation time-step as the time unit. Note that, 
since we are concerned with incompressible flows, we do not need to introduce 
a dimension of mass.
The relaxation time is chosen so that the kinematic viscosity
is $\nu=0.1$ in dimensionless units. 
The pressure gradient is $\nabla p= 6.25~10^{-6}$,
yielding a mean velocity in the range $ 2.0~10^{-4} < U_x < 2.0~10^{-2}$,
for separation between surfaces of the fracture between 
$8 \leq h \leq 64$. Then, for gap sizes $h$ not larger than
$h_{S}=20$, the Reynolds number is $\text{Re} = U_x h / \nu < 1$ and the
flow is governed by the Stokes equations, which are invariant
under a velocity rescaling.

\section{Permeability of self-affine fractures}
\label{permeability}
Let us consider first the case in which the two matching surfaces of
a fractured rock are separated by a distance $h$ in the normal direction,
with no lateral shift between the two complimentary surfaces of the fracture. 
In this situation, the aperture of the fracture is clearly constant everywhere, 
but the effective local aperture for fluid flow, i.e., the local width
of the fracture channel normal to the mean flow direction strongly
depends on the local angle between the surface and its mean plane. 
The following theoretical analysis will be based on a kind of local 
lubrication approximation, in which the fracture is divided into
smaller rectangular blocks. The linear size $\xi$ of the blocks will be estimated
as the length scale over which fluctuations in the surface height
are small compared to the distance between surfaces and thus, 
each of these basic blocks can be considered, approximately, as two facing
planes separated a distance $h$ and with varying orientation angles
with respect to the mean plane of the fracture.

\subsection{Analogue resistor network}
\label{resistor}

The transport properties of these narrow fractures can be modeled
using effective-medium ideas. The representation of the fracture 
in terms of quasi-linear blocks with random orientation
can be mapped onto a regular two-dimensional square lattice,
of lattice spacing $\xi$, where the disorder only enters the 
distribution of hydraulic conductances lying on
each of the lattice bonds. Moreover, fluid flow is a locally conserved 
quantity at the lattice nodes and therefore, as we shall discuss in this section,
this representation of the fracture is completely analogous to the classical 
random resistor networks that model electrical transport in disordered 
media \cite{kirkpatrick73,koplik90}. 

Consider the two opposite ends $a$ and $b$ of a single quasi-linear
block of a fracture, and its representation by a conductance joining the
corresponding two lattice nodes $a$ and $b$. Moreover, consider that a pressure drop
is imposed between the two nodes, and that the lattice-bond joining $a$ and $b$
is oriented either parallel or perpendicular to the mean flow.
Using the Poiseuille result for the flow through this segment of the fracture
we have that
\begin{equation}
\label{block}
Q_{a \to b} = - 
\frac{\left[ h \cos(\theta) \right]^3}{12 \mu}
\frac{\Delta p_{ab}}{[\xi / \cos(\theta)]}
\end{equation} 
where $Q_{a \to b}$ is the flow rate from node $a$ to node $b$,
$\xi$ is the linear size of the block, and $\theta$ is the orientation 
angle of the block, with respect to the mean plane of the fracture, in the 
direction of the imposed pressure drop $\Delta p_{ab}$. Let us note, that a 
slightly different approximation to the transport of fluid in a single block 
can be made by modeling each block as a cylinder of diameter $h \cos(\theta)$
and length $\xi$. However, these two approximations will only differ by a numerical factor,
given that the ratio of the permeability of a cylinder to that
of two parallel planes is independent of the separation between surfaces, 
$k_{straight}/k_{cylinder}=32/12$.

We can rewrite the previous equation in terms of the
bond conductance $g_{ab}$ as
\begin{equation}
\label{gab}
\Delta p_{ab} = - g^{-1}_{ab} Q_{a \to b} 
\end{equation}
where $g_{ab}$ is a random variable related to the
actual height distribution of the fracture surface
through the orientation angle $\theta$.

Furthermore, the fluid is incompressible and flux
is conserved at each lattice node,
\begin{equation}
\label{kirchof}
\sum_{a \rightarrow b} Q_{a \to b} = 0
\end{equation}
where the sum is over all nodes $a$ connected
directly to $b$. The previous two equations
are formally equivalent to the equations
of an electrical-resistor network with the
correspondence \cite{koplik82}
\begin{eqnarray}
\label{analogy}
{\rm pressure} &\Leftrightarrow& {\rm voltage} \\ \nonumber
{\rm fluid~flux} &\Leftrightarrow& {\rm electric~current} \\ \nonumber
{\rm hydrodynamic~conductance} &\Leftrightarrow& 
{\rm electrical~conductance}
\end{eqnarray}

Replacing Eq.~(\ref{block}) into Eq.~(\ref{kirchof}) 
we obtain the following system of
linear equations for the pressure at each node of
the network,
\begin{equation}
\sum_{a \rightarrow b} g_{ab} (p_a-p_b) = 0
\end{equation}

Then, the problem is now to solve the previous system of equations
in the case where the hydrodynamic conductances $g_{ab}$ vary 
according to some probability distribution, and we shall discuss now 
an approximation commonly used to solve this type of problems 
in the physics of disordered media, namely {\it effective-medium theory}.

\subsection{Effective-medium approximation}
\label{EMA}

The effective-medium approximation (EMA) is a standard method 
commonly used to calculate effective properties of a microscopically disordered 
medium, in which the random microscopic parameters are replaced by 
a certain mean value, chosen in a self-consistent way 
(a detailed discussion of this approximation can be found eslewhere
\cite{kirkpatrick73,sahimi93}).

We shall summarize here the main results obtained within 
EMA for random-resistor networks but adapted to flow networks, 
using the correspondence given in (\ref{analogy}), and we shall 
make use of these results to compute the permeability of the fracture
in Sec.~\ref{pdf} and the magnitude of the velocity fluctuations
in Sec.~\ref{fluctuations}.

\begin{figure}[htb]
\includegraphics[angle=-90,width=\W]{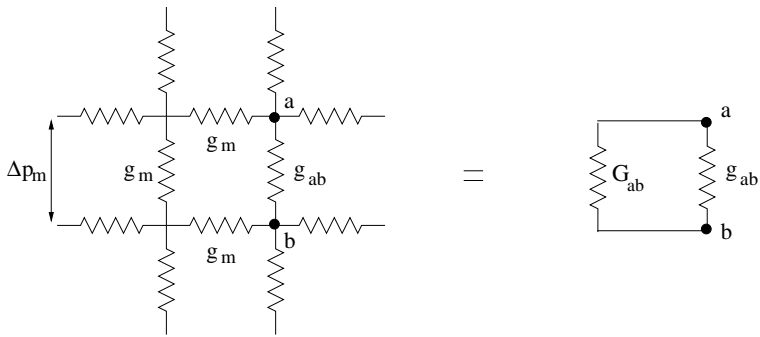}
\caption[network]{\label{network} Representation used to determine the voltage across
a conductance $g_{ab}$ immersed in a uniform network of
conductances $g_m$, where $g_m$ is the mean EMA conductance.}
\end{figure}

Suppose we have an infinite resistor network, with a coordination number $z=4$
and individual bond conductances $g_{ab}$, as the one represented in 
Fig.~\ref{network} (the general case where $z$ can take any integer value is 
completely analogous and is discussed elsewhere \cite{koplik82,kirkpatrick73}).
The idea then is to choose a mean EMA hydraulic conductance $g_m$,
in a self-consistent way, such that it reproduces the average local field. 
That is, we require that the extra pressure differences induced
when one individual conductance reverts from $g_m$ to its original
value $g_{ab}$ average to zero.
In that case, when one bond conductance, oriented along the external
pressure drop and surrounded by the effective medium, is replaced by 
an individual random conductance having the value $g_{ab}$, 
the uniform field solution, in which the pressure drops
a constant amount $\Delta p_m$ across any bond conductance oriented
along the imposed pressure difference, fails to satisfy fluid flux 
conservation. 
In fact, there will be an excess of fluid flux in that bond of
\begin{equation}
\label{current}
\Delta Q = (g_m-g_{ab}) \, \Delta p_m
\end{equation}
This excess in fluid flux corresponds to an extra pressure drop
$\delta p_{ab}$, which can be calculated 
in terms of the conductance $G_{ab}$ of 
the network between points $a$ and $b$ in the
absence of $g_{ab}$. After some straightforward 
calculations, the induced pressure drop across $ab$
results \cite{kirkpatrick73,koplik82},
\begin{equation}
\delta p_{ab} = \Delta p_m \frac{g_m-g_{ab}}{g_m+g_{ab}}.
\end{equation}   

The value of the individual bond conductance $g_{ab}$ and hence 
that of the induced pressure difference $\delta p_{ab}$ is a 
random variable and, the criterion to choose $g_m$ within
EMA is to require that the average of $\delta p_{ab}$
with respect to the probability distribution of local 
conductances $g_{ab}$ vanish:
\begin{equation}
\label{ema}
\left\langle \frac{g_m-g}{g_m + g} \right\rangle = 0
\end{equation}

A detailed study of circumstances under which the
EMA is expected to be valid is presented in Ref.~\cite{koplik81}. 

\subsection{Probability distribution of conductances and mean EMA conductance}
\label{pdf}

As discussed in the previous section, in order to determine the mean EMA 
conductance $g_m$, we first need to obtain the probability distribution of 
the individual bond conductances $g_{ab}$.
The conductance of a single block is determined by the angle 
between the local orientation of the surface and the mean plane 
of the fracture (see Eq.~(\ref{gab})). In terms of the height 
difference $Z$ between the two ends of the block, 
the bond conductance can be written as,
\begin{equation}
\label{gZ}
g(Z) = G_0 \left( \frac{\xi^2}{\xi^2+Z^2} \right)^2, 
\qquad G_0 = \frac{h^3}{12 \mu \xi}.
\end{equation}
Then, the average over local conductances
given in Eq.~(\ref{ema}) can be performed in terms 
of the distribution of height differences $Z$, between
points of the fracture surface separated a distance 
$\xi$ (note that $\xi$ is the distance between two
points lying on the fracture surface projected over 
the mean plane, i.e. the lattice constant.) 

Experimental measurements indicates that the distribution of heights 
$p(Z)$ can be accurately described by a Gaussian distribution, at least for
low-order moments \cite{plouraboue95,mourzenko96}. 
Guided by these results, the numerical procedure used to
generate the fracture surfaces is intended to produce self-affine surfaces with Gaussian
fluctuations of the height, as discussed in Sec.~\ref{numericalsurfaces}. 
Moreover, the perturbative analysis we shall present 
shows that higher moments of the height distribution function 
are related to higher order corrections to $g_m$ and,
therefore, the exact functional form
of the distribution is not relevant but only its low-order moments.

Finally, assuming that neighboring conductances are uncorrelated, 
we can rewrite the integral equation (\ref{gab}), which
defines the EMA conductance $g_m$, in terms of $p(Z)$,
\begin{equation}
\label{inteq}
\int dZ p(Z) \frac{g_m (1+ (Z/\xi)^2)^2 - G_0}{g_m (1 + (Z/\xi)^2)^2 + G_0 } = 0
\end{equation} 
where
\begin{equation}
\label{gaussian}
p(Z) = \frac{1}{\sqrt{2\pi \sigma^2_z(\xi)}}
\exp{\left(-\frac{Z^2}{2 \sigma^2_z(\xi)}\right)}.
\end{equation}
In terms of the reduced variable $x=Z/\sigma_z(\xi)$, which is normally
distributed, we obtain,
\begin{equation}
\label{intend}
\int dx \frac{1}{\sqrt{2\pi}} \exp{\left(-\frac{x^2}{2}\right)}
\frac{g_m (1+\epsilon^2 x^2)^2 - G_0}{g_m(1+ \epsilon^2 x^2)^2+ G_0 } = 0,
\end{equation} 
where the dimensionless parameter $\epsilon = \sigma_z(\xi)/\xi$ measures the
ratio between the average magnitude of the fluctuations in the height of the 
surface $\sigma_z(\xi)$ over a single block, and the size of these blocks 
$\xi$. In general, if the individual blocks of the fracture are smaller than the 
characteristic length $\xi \ll \ell$ we have that $\epsilon < 1$, 
and $\epsilon > 1$ for $\xi$ larger than $\ell$. However, since we consider
the case in which the characteristic length is small compared to the simulated
length scales ($\ell \sim 10^{-3} \ll 1$, see Sec.~\ref{topo}), we have that,
\begin{equation}
\epsilon \equiv 
\frac{\sigma_z(\xi)}{\xi} 
= \left( \frac{\xi}{\ell} \right)^{\zeta-1} \ll 1.
\end{equation}
Therefore, we can then evaluate the mean EMA conductance in a perturbative weak-disorder 
expansion by computing successive terms in the series expansion of $g_m$ in the 
perturbative parameter $\epsilon$,
\begin{equation}
g_m = g_m^0 + g_m^1 \epsilon^2 + g_m^2 \epsilon^4 + \dots
\end{equation}
Replacing this series expansion of $g_m$ into Eq.~(\ref{intend}),
and rearranging terms by their order in $\epsilon$, we obtain,
\begin{eqnarray}
\nonumber
&& \, \epsilon^0 \Bigg\{\frac{ g_m^0 - G_0}{ g_m^0 + G_0} \Bigg\} \\ \nonumber 
&+& \, \epsilon^2 \Bigg\{ \frac{2G_0}{( g_m^0+G_0)^2} 
\left[ g_m^1+2 \left\langle x^2 \right\rangle g_m^0\right] \Bigg\}
\\  \nonumber
&+& \, \epsilon^4 \Bigg\{ \frac{2G_0}{( g_m^0+G_0)^3} 
\left[ \left[ g_m^2 (g_m^0+G_0) - \left( g_m^1 \right)^2 \right] \right. \\ \nonumber
&& + \left. \left[ 2 g_m^1 (G_0-g_m^0) \right] \left\langle x^2 \right\rangle
+ \left[ g_m^0 (G_0 - 3 g_m^0)  \right] \left\langle x^4 \right\rangle
\right] \Bigg\} \\ 
&+& O(\epsilon^6) = 0.
\end{eqnarray}
Furthermore, since the previous equation must be satisfied for any value of $\epsilon$,
then each term in the previous equation should be exactly zero, yielding the following
expressions for $g_m^i$, 
\begin{subequations}
\label{expansion}
\begin{eqnarray}
g_m^0 &=& G_0, \\ 
g_m^1 &=& -2 \left\langle x^2 \right\rangle \, G_0, \\
g_m^2 &=& \left[ \left\langle x^4 \right\rangle + 
2 \left\langle x^2 \right\rangle^2 \right] \, G_0.
\end{eqnarray}
\end{subequations}
Finally, computing the second and fourth moments of the  Gaussian distribution given 
by Eq.~(\ref{gaussian}) we obtain the expansion in power series of $\epsilon$
for the mean EMA conductance,
\begin{equation}
\label{gmEMA}
g_m = G_0 \left[ 1 - 2 \epsilon^2 + 5 \epsilon^4 + O(\epsilon^6) \right].
\end{equation}
Let us remark that only the second and fourth moments of the Gaussian probability 
distribution of heights given in Eq.~(\ref{gaussian}) were involved in the previous 
calculation, and higher moments only occur in higher order terms, $O(\epsilon^6)$.
Thus, as already mentioned, this validates the use of a Gaussian distribution of heights, 
which was found to accurately describe experimental measurements for low-order moments.
On the other hand, to compute higher order terms in the $\epsilon$ expansion of the 
electrical conductivity of the network would require a different approach, e. g., 
a perturbative weak-disorder expansion of the conductivity in terms of the
moments of the probability distribution of bond conductances to directly solve 
Eq.~(\ref{kirchof}), since EMA would no longer be an accurate approximation \cite{luck91}.

\begin{figure}[htb]
\includegraphics[angle=-90,width=\W]{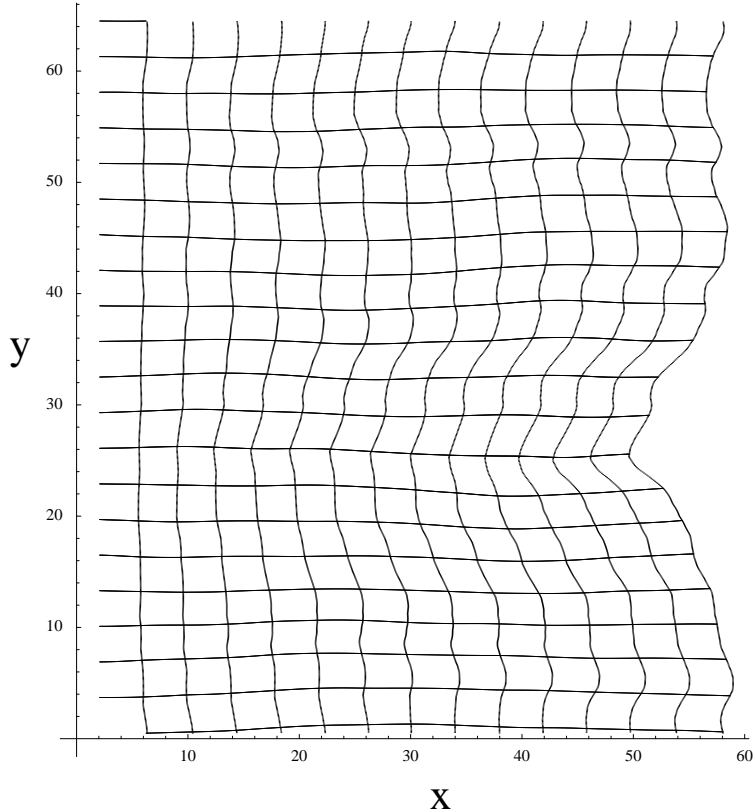}
\caption[front]{\label{front} Streamlines and front propagation in the gap-averaged
flow field. The imposed pressure drop and thus the mean flow are along the 
$x$ direction. The corresponding fracture surface is shown in Fig.~\ref{surface}.
The system size is $L=64$ and the gap between fracture surfaces is $h=8$.}
\end{figure}

The previous result for the mean EMA conductance $g_m$ shows an important
feature of the fractures, in that 
the same result, up to second order in $\epsilon$, is 
obtained if those conductances that are perpendicular 
to the mean flow are eliminated from the resistor network, 
i.e. it predicts a {\it quasi-one-dimensional} flow. The two-dimensional 
character of the network only affects higher order terms, $O(\epsilon^4)$.
In fact, deriving from the previous equation the permeability
of the fracture we obtain,
\begin{equation}
\label{perm}
k \approx \frac{h^2}{12} 
\left[ 1-2 \left( \frac{\sigma_z(\xi)}{\xi} \right)^2  \right],
\end{equation}
which is the same result as obtained in Ref. \cite{GK1}
for two-dimensional fractures (see Eq.~(29) in \cite{GK1}).

In Fig.~\ref{front} we present the streamlines and the propagation of an initially
flat front in the velocity field inside a self-affine fracture, where the flow field
was averaged over the gap of the fracture. That is, velocities along and perpendicular
to the mean flow direction are averaged in the $z$ direction over the aperture of 
the fracture, rendering a two-dimensional flow field $\vec{u}(x,y)$. 
Then, using this flow field, the streamlines and the propagation of
a flat front initially located at $x=0$ are computed.
The pressure gradient, and therefore the mean flow velocity, are along 
the $x$ direction, $\langle \vec{u}(x,y) \rangle = U_x \, \check{x}$. 
It can be seen that, as predicted by the effective-medium
approximation, lateral fluctuations in velocity are small
compared to the mean flow, and streamlines are approximately
straight lines oriented along the direction of the imposed pressure difference.
However, there is a substantial difference with the 2d simulations, in that in the 
2d case the gap-averaged mean flow velocity is constant due to flux conservation, 
whereas in the 3d case,
as is clear from Fig.~\ref{front}, fluctuations in the gap-averaged
velocity occur, and result in the broadening of the initially flat front.

\subsection{Permeability dependence on the fracture gap}
\label{KvsH}

Thus far we have obtained the relation between the size of the quasi-linear blocks 
composing our model fracture and the mean EMA conductance of the system $g_m$. To 
further obtain the dependence of $g_m$ on the size of the fracture gap, it is first 
necessary to estimate the size of the unit blocks $\xi$ in terms of $h$. The linear 
size of the unit blocks was defined as the characteristic length over which the channel 
formed by the two fracture surfaces can be considered, in what the flow is concerned, 
as a straight one. Such an approximation is valid when the vertical fluctuations of the
fracture surface over the characteristic size of a unit block are a small fraction of $h$, 
\begin{equation}
\label{quasilinear}
\frac{\sigma(\xi)}{h} = C_{\xi} \ll 1,
\end{equation}
where $C_{\xi}$ shall be treated as a fitting parameter to our numerical 
simulations and consitency with the previous requirement ($C_{\xi} \ll 1$) 
will be shown.

In view of Eq.~(\ref{variance}) yields,
\begin{equation}
\label{xi}
\xi \simeq \ell \left[ C_{\xi} \frac{h}{\ell} \right]^{1/\zeta}.
\end{equation}
Thus, the small parameter $\epsilon$ is, in terms of $h$,
\begin{equation}
\label{eh}
\epsilon \simeq \left[ C_{\xi} \frac{h}{\ell} \right]^{\frac{\zeta-1}{\zeta}}
\end{equation}
Then, replacing the previous result into Eq.~(\ref{perm}) 
we obtain the dependence of the permeability
of a three-dimensional self-affine fracture in terms of the 
separation between surfaces,
\begin{equation}
\label{k}
k = \frac{h^2}{12} 
\left[ 1 - 2 \left[ C_{\xi}  \frac{h}{\ell}\right]^{2\frac{\zeta-1}{\zeta}}
\hspace{-0.4cm} +5 \left[ C_{\xi} \frac{h}{\ell}\right]^{4\frac{\zeta-1}{\zeta}} \hspace{-0.4cm} 
+ O\left( \left[C_{\xi}  \frac{h}{\ell}\right]^{6\frac{\zeta-1}{\zeta}}\right)
\right]
\end{equation}
As mentioned earlier, this result is the same, up to second order in $\epsilon$, as the 
one obtained for two-dimensional fractures (see Eq.~(33) in \cite{GK1}).  In order to 
test this result numerically, we first recast it in terms of the flow rate per unit width 
$q$. In a straight channel of height $h$, length $L$, with fixed pressure drop $P$, 
the flow rate per unit width is $q_0= h^3 P / 12 \mu L$. Thus, in view of Eq.~(\ref{k}), 
the deviation from $q_0$ in a fracture of gap size $h$ is given by,
\begin{equation} 
q_0 - q \approx \, \left[ \frac{1}{6 \mu} \frac{P}{L} 
\left(\frac{C_{\xi}}{\ell}\right)^{\frac{2-2\zeta}{\zeta}} 
\right] \, h^{\frac{5\zeta-2}{\zeta}}
\label{flow-disp}
\end{equation}
where $C_{\xi}$ is an adjustable parameter expected to be small.

\begin{figure}
\includegraphics[angle=-90,width=\W]{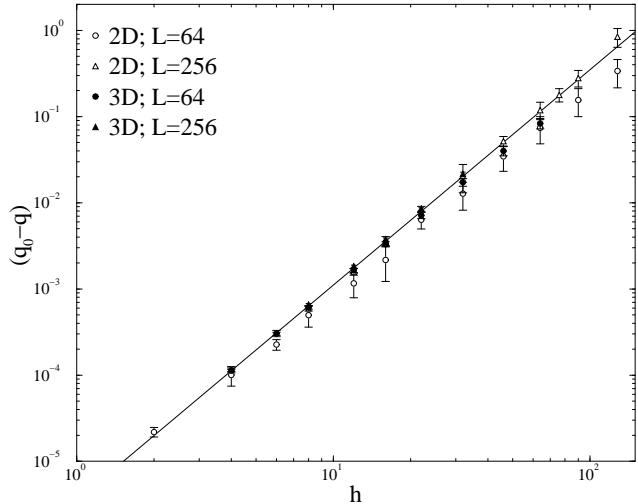}
\caption[flow3D2D]{\label{flow3D2D} Deviation from the flow rate per unit width
in a straight channel $q_0-q$ as a function of the size of the vertical gap $h$. 
We compare results from 2d and 3d simulations and from two different sizes $L$
of the system, $L=64$ and $L=256$.
The solid line is a fit to the numerical results obtained in the 
largest 2d system.}
\end{figure}

In Fig.~\ref{flow3D2D} we compare numerical results, of the deviation from the flow
rate per unit width in straight channels $q_0$, obtained in $q-q_0$ in 2d and 3d fractures. 
In both cases, the fracture surfaces (fracture curves in the 2d case) have the same 
roughness exponent and the same characteristic length $\ell$, hence, both have,
in average, the same amplitude of the vertical fluctuations in surface height. 
We also show in Fig.~\ref{flow3D2D} the scaling relation given by Eq.~(\ref{flow-disp}) 
with the value of $C_{\xi}$ obtained for two-dimensional fractures in Ref.~\cite{GK1},
 $C_{\xi}\simeq 0.1$, which as expected is a small number consistent with
the approximation of quasi-linear blocks discussed before and 
given by Eq.~\ref{quasilinear}.  It can also be seen that, in agreement with the 
effective medium prediction, both 2d and 3d results are very similar. 
As discussed in Sec.~\ref{pdf} the difference 
between the 2d and 3d cases comes in the fourth order term, $O(\epsilon^4)$,
and its correction is of relative magnitude $5 \epsilon^2 /2$. For the results presented 
in Fig.~\ref{flow3D2D} the largest value of $\epsilon$ is $\epsilon \sim 0.15$ 
(corresponding to the smallest gap size $h=4$), hence the relative magnitude 
of the fourth order term is $5 \epsilon^2 /2 \sim 0.05$, which 
is consistent with the observed similarity between the results obtained in 2d and 3d.

\subsection{Velocity fluctuations}
\label{fluctuations}

\begin{figure}
\includegraphics[angle=-90,width=\W]{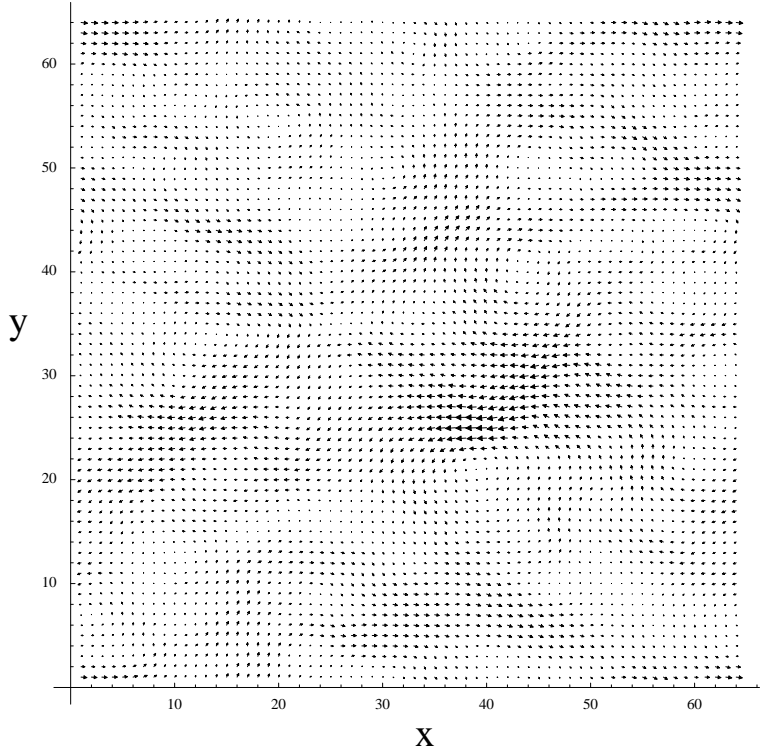}
\caption[field]{\label{field} Fluctuations in the gap-averaged velocity field
The fracture surface used in the numerical simulation is shown in
Fig.~\ref{surface}, and the corresponding streamlines and front propagation
are presented in Fig.~\ref{front}. The size of the system is $L=64$
and the gap size is $h=8$.}
\end{figure}

In the previous section we showed that the flow rate per unit width
in 2d and 3d fractures, which have the same characteristic length $\ell$, 
is very similar and, in fact, both have the same
scaling behavior as a function of the gap size $h$.
However, as was discussed in Sec.~\ref{pdf}, there is an important feature
of the flow field in 3d fractures that is not present in 
two-dimensional simulations, that is the presence of velocity 
fluctuations even after the fluid velocity is averaged over the gap of 
the fracture. In Fig.~\ref{field} we present the 
fluctuations in the gap-averaged velocity, obtained by means of the 
LBM in a fracture of size $L=64$ and gap size $h=8$. The corresponding 
fracture surface is shown in Fig.~\ref{surface} and the streamlines corresponding
to the gap-averaged two-dimensional velocity field are
plotted in Fig.~\ref{front}.

Let us then investigate the magnitude of these fluctuations in the 
gap-averaged velocity in the direction of the mean flow, i. e. 
$\delta u_x = u_x(x,y) - U_x$. Previously, in Sec.~\ref{EMA},
we calculated the mean EMA conductance $g_m$, in a self-consistent way, 
requiring that the extra pressure drop $\delta p_{ab}$, induced between lattice nodes
$a$ and $b$ when a bond conductance $g_m$, oriented in the direction of the external 
pressure drop, is replaced by a randomly chosen conductance $g_{ab}$, should average to zero
(see Eq.~(\ref{ema})). In a similar way, we can estimate the fluctuations in the 
gap-averaged velocity field in the direction of the mean flow, in terms of the mean variance 
of these extra pressure drops. In fact, we might think of the random extra-pressure-drops
$\delta p_{ab}$ as the source of the velocity fluctuations.
Let us mention, that the same result would be obtained if
the fluctuations are computed in terms of the mean variance of the 
induced excess of fluid flux from Eq.~(\ref{current}).

The variance of the induced extra pressure drops $\delta p_{ab}$ across 
a random conductance bond $g_{ab}$ oriented in the direction of the flow,
is given by,
\begin{equation}
\left\langle \delta p^2 \right\rangle = 
\left(\Delta p_m\right)^2 \left\langle \left[ \frac{g_m-g}{g_m+g} \right]^2 \right\rangle.
\end{equation}

Repeating now the procedure we followed to calculate the leading terms of $g_m$,
i. e. Eqs.~(\ref{inteq}), (\ref{gaussian}) and (\ref{intend}), and replacing $g_m$ by
its series expansion in powers of $\epsilon$ given in Eq.~(\ref{expansion}), we obtain
\begin{equation}
\label{dv}
\left\langle \delta p^2 \right\rangle = 
\epsilon^4 \left(\Delta p_m \right)^2
\left\langle \left[ x^2-\left\langle x^2 \right\rangle \right]^2 \right\rangle.
\end{equation}
The previous is a general result, in that it is independent of the particular distribution
of heights of the fracture surface $p(x=Z/\sigma_z(\xi))$. 
Now, if we replace in this equation the second and fourth
moments of the distribution by their normal values, $\langle x^2\rangle=1$ and 
$\langle x^4\rangle=3$, we obtain,
\begin{equation}
\label{dp}
\left\langle \delta p^2 \right\rangle = 
2 \epsilon^4 \left(\Delta p_m \right)^2.
\end{equation}
Finally, using the fact that the relative fluctuations in the pressure drop 
are equal to the relative fluctuation in the velocity,
$\delta p / \Delta p_m = \delta u_x / U_x$, we obtain
the variance in the gap-averaged velocity $u_x$ normalized by its mean value,
\begin{equation}
\label{sigmaU}
\delta^2 \equiv \frac{\sigma^2_u}{U_x^2}=\frac{\left\langle \delta u_x^2 \right\rangle}{U_x^2} =  
\frac{\left\langle \delta p^2 \right\rangle}{\left(\Delta p_m \right)^2}=
2 \, \epsilon^4.
\end{equation}

\begin{figure}
\includegraphics[angle=-90,width=\W]{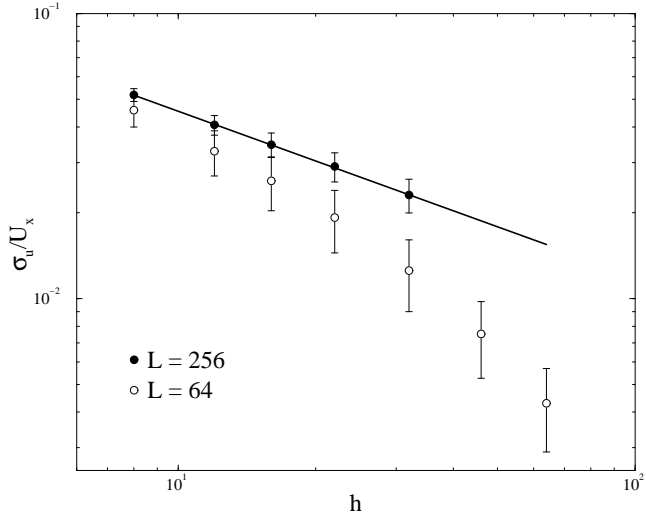}
\caption[ux]{\label{ux} Relative magnitude of the mean velocity fluctuations
in the direction of the flow, as a function of the gap size $h$. 
Two different sizes of the fracture $L=64$ and $L=256$ are
presented. Solid line is a fit to the numerical results obtained
in the largest system ($L=256$).}
\end{figure}

In order to test this result numerically, let us
rewrite the previous equation in terms of the gap size $h$,
\begin{equation}
\delta =  C_f \sqrt{2} 
\left( \frac{h}{\ell} \right)^{(2\zeta-2)/\zeta} 
\end{equation}
where again $C_f$ is an adjustable parameter.

In Fig.~\ref{ux} we show the numerical results obtained for the normalized fluctuations 
of the gap-averaged velocity in the direction of the mean flow, as a function of the 
distance $h$ between unshifted fracture surfaces. We compare results obtained for two 
different sizes of the system $L=64$ and $L=256$. We find a good agreement with the 
predicted exponent for the largest system. The fitted exponent is $-0.58 \pm 0.08$
and the predicted one is $(2-2\zeta)/\zeta=-0.5$. The adjustable parameter is found 
to be $C_f=2.1 \pm 0.9$. Note that $C_{\xi}$ can be computed from $C_f$ by means of
Eq.~\ref{eh},  obtaining $C_{\xi}\simeq0.2$, which is similar to our 
previous determination and is also consistent with the quasi-linear approximation
given by Eq.~\ref{quasilinear}. On the other hand, 
the scaling behavior of the smaller system deviates from the predicted one. We believe 
that this discrepancy is due to correlations in the velocity fluctuations, and we shall 
show in Sec.~\ref{autocorrelation} that the correlation length in the velocity fluctuations is,
in this case, of the order of the size of the system and hence velocity fluctuations are 
significantly reduced. Moreover, we shall also show, in Sec.~\ref{D_vs_h}, that as the 
separation between surfaces increases the correlation length also increases, and the mentioned 
finite-size effect becomes more important. This fact explains why the larger deviations from 
the expected scaling behavior observed in Fig.~\ref{ux} occur at larger separations 
between surfaces. Let us note, that the correlation length of the fluctuations in 
the gap-averaged velocity is not necessarily equal to the size $\xi$ of the quasi-linear 
blocks. In fact, in the two-dimensional case there are no fluctuations in the gap-averaged velocity
and still the description in terms of quasi-linear blocks was shown to be valid \cite{GK1}.

\begin{figure}
\includegraphics[angle=-90,width=\W]{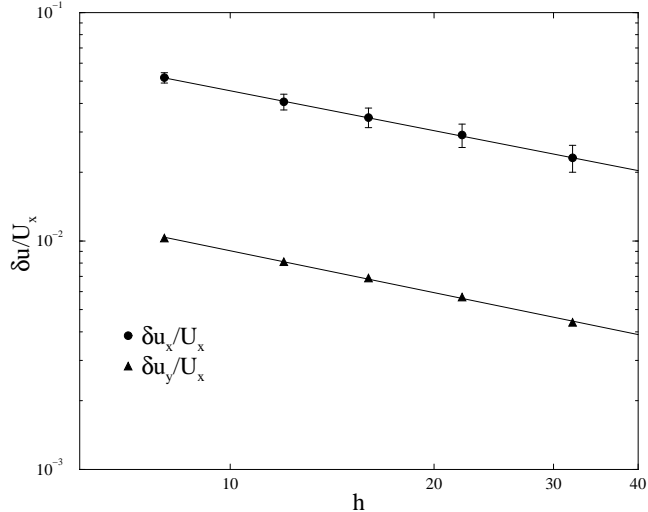}
\caption[uxy]{\label{uxy} Relative magnitude of the mean velocity fluctuations,
both parallel and perpendicular to the mean flow, as a function
of the vertical separation between surfaces $h$. The mean flow is
in $x$. The size of the system is $L=256$.}
\end{figure}

In Fig.~\ref{uxy} we compare the fluctuations in the gap-averaged velocity in both directions, 
along and perpendicular to the mean flow. It can be observed that the scaling of the fluctuations in
the velocity perpendicular to the mean flow is also given by Eq.~(\ref{ux}), with a fitted exponent
$-0.6\pm0.8$. The large uncertainty in the fitted exponent comes from the fact that the 
fluctuations perpendicular to the mean flow have zero mean value and its variance presents large
variations between different fractures. Finally, it can be observed that fluctuations perpendicular 
to the mean flow are weaker than fluctuations parallel to the flow, 
$\delta u_x \sim 5 \, \delta u_y$. The approximately constant ratio between fluctuations along 
and normal to the mean flow, $\delta u_x/\delta u_y \sim 5$, is consistent with the conservation
of fluid flux. On the other hand, the magnitude of the velocity fluctuations are clearly related 
to the spatial correlations in the velocity field, and we will find an analogous 
asymmetry in the autocorrelation function of the velocity
fluctuations. Let us mention that similar results are obtained within
a macroscopic-continuum approach to the problem of transport in heterogeneous
porous formations, where the magnitude of the fluctuations in the direction
of the mean flow are found to be three times larger than the 
perpendicular ones in two-dimensional systems \cite{dagan84}

\subsection{Fractures with shifted surfaces}
\label{shift}

Usually, when a rock is fractured its two matching surfaces
are laterally shifted, that is the displacement between 
them is not only vertical but also parallel to the mean plane of the 
fracture. We shall now  consider this situation, in which the upper surface
of the fracture is laterally shifted by a vector $\vec{d}=(d_{\parallel}, d_{\perp})$ 
lying in the mean plane of the fracture. However, we will only consider the
case where the fracture is distinctly open, that is the two surfaces do not touch each 
other at any point. In this the aperture of the fracture is no longer constant
but becomes a random function of the position $a_d(x,y) = z(x + d_x,y + d_y)-z(x,y)+h$. 
Let us consider first, the case in which the lateral shift lies in the direction of the 
mean flow, i.e. $d = d_{\parallel}$. In Ref. \cite{GK1} we investigated how such a 
lateral shift modifies the permeability of two-dimensional fractures. We showed that, 
for large separation between surfaces, such that the characteristic size $\xi$ of the 
quasi-linear blocks is much larger than the lateral shift $d_{\parallel} \ll \xi$, 
there is little change in the fracture geometry compared to the unshifted case, and that 
the permeability is asymptotically the same. On the other hand, when surfaces nearly touch 
at some point, the permeability will be dominated by the large pressure drop around this point, 
as the fluid is constrained to flow through this narrow gap. Thus, as the surfaces become closer, we
found a decrease in the permeability as compared to the unshifted case. The case of 
three-dimensional fractures is somewhat different. For large separation between surfaces 
($d_{\parallel} \ll \xi$; Recall that $\xi = \xi(h) \propto h^{1/\zeta}$) a behavior similar 
to the unshifted case is again expected, since the change in the geometry of the fracture is 
asymptotically negligible, and therefore, the scaling relation given by Eq.~(\ref{flow-disp}) 
should still apply. On the hand, when surfaces are close to each other the fluid is no 
longer forced to flow through the narrow gaps, where the minimum separation between surfaces 
occur, as in the 2d case. In three dimension, the fluid can avoid this low permeability regions
by flowing around them. Thus, we might expect that the fluid rate per unit width $q$ 
is bounded by the behavior in two-dimensional fractures, that is the upper-bound of $q$ given 
by the two-dimensional flow rate in the case without lateral shift, and the lower-bound given 
by the flow rate per unit width in two-dimensional fractures with the same lateral shift, 
$q^{2d}_{d=d_{\parallel}}<q<q^{2d}_{d=0}$. 

\begin{figure}
\includegraphics[angle=-90,width=\W]{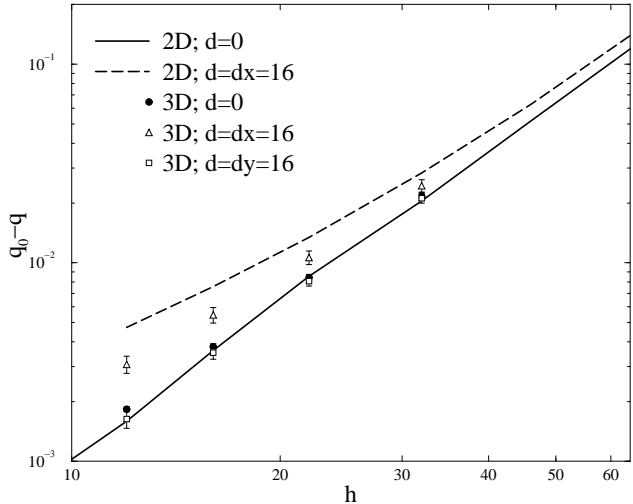}
\caption[shift3D]{\label{shift3D} Deviation of the flow rate per unit width 
from a straight channel, as a function of the mean vertical separation 
between surfaces $h$. The solid line corresponds to a 2d fracture formed by 
two complimentary surfaces that are simply displaced in the vertical direction, 
while the dashed line corresponds to results obtained when the upper surface 
of the 2d fracture is also shifted in the direction of the mean flow $dx=16$. 
Solid circles correspond to 3d simulations with no lateral
shift between the matching surfaces. Open triangles and squares correspond to 
the upper surface shifted in the direction of the flow $d_{\parallel}=16$ and 
perpendicular to it $d_{\perp}=16$, respectively.}
\end{figure}

In Fig.~\ref{shift3D} we present the correction to the flow rate per unit width,
$q_0-q$, as a function of the distance between surfaces, $h$.
In agreement with our previous discussion, the correction to the
flow rate obtained for three-dimensional fractures lies above the
2d case without lateral shift (upper bound for the total flow rate $q$) 
and below the two-dimensional results obtained for the same 
lateral shift $d=d_{\parallel}$ (lower bound for the total flow rate $q$). 
It can also be observed that, as we discussed earlier, for large separations 
between surfaces all the results converge to the unshifted situation
described by Eq.~(\ref{flow-disp}).

\begin{figure}
\includegraphics[angle=-90,width=\W]{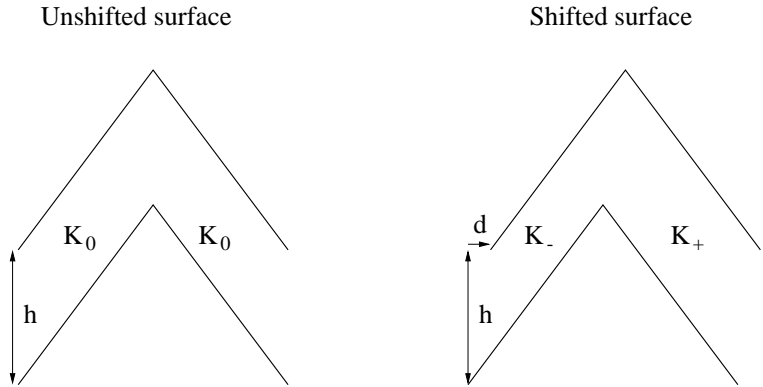}
\caption[perm]{\label{perm_view}
Schematic representation of the intersection of a small region
of the fracture with a plane containing both the displacement vector
$\vec{d}$ and the normal to the fracture surfaces. 
The depicted region contains two consecutive quasi-linear blocks
of size $\xi$ in the unshifted and shifted cases. The local
permeabilities are assumed equal in the unshifted case $k=k_0$.}
\end{figure}

In three dimensions, it is also possible to have a displacement 
perpendicular to the direction of the flow, $d=d_{\perp}$. 
Let us then investigate how the orientation of the lateral shift
affects the permeability of the fracture. In Fig.~\ref{perm_view} 
we show a schematic representation of the intersection of a small region of
the fracture, approximated by two consecutive quasi-linear blocks of
size $\xi$, with the plane that contains both the displacement vector $\vec{d}$ 
and the vector normal to the mean plane of the fracture. Although extremely simplified, this 
schematic representation of the fracture shows the effect of the shift 
on the local permeability. It can be seen that, upon a lateral displacement,
the unshifted local permeability of a single linear block, $k_0$,
decreases, $k_-$, or increases, $k_+$, depending on the orientation of the 
quasi-linear block. Specifically, an increase (decrease) in the permeability 
corresponds to $\theta < 0 ~ (>0)$. 
Furthermore, when the shift is in the direction of the flow the two channels 
shown in Fig.~\ref{perm_view} will be in {\it series}, i.e. 
approximately preserving the fluid flux. On the other hand, if the shift is in the
direction perpendicular to the mean flow the two blocks will be in {\it parallel},
that is having approximately the same end-to-end pressure drop. It can be shown that,
this somehow naive model predicts a reduction in the permeability upon a shift in the 
direction of the flow, and a smaller correction in the case
where the shift is perpendicular to the flow, in agreement with the
results presented in Fig.~\ref{shift3D}. The same effect can be observed 
in the experimental work reported in Ref. \cite{auradou01}, where the flow rate was 
observed to be larger in the direction perpendicular to the shift between the surfaces 
(see Fig.~4 in \cite{auradou01}).  It can also be observed in the experimental
work presented in Ref.~\cite{auradou01} that the difference
between the flow perpendicular and parallel to the shift decreases
as the separation between the surfaces is increased, which is also
in agreement with our results (see Fig.~5 in \cite{auradou01}).
Let us mention that, in our simulations, the largest contrast between 
$q_{\perp}$ and $q_{\parallel}$, obtained for the smallest gap size $h=12$, is found to be
$q_{\perp}/q_{\parallel} \sim 1.24$.

\begin{figure}
\includegraphics[width=\W]{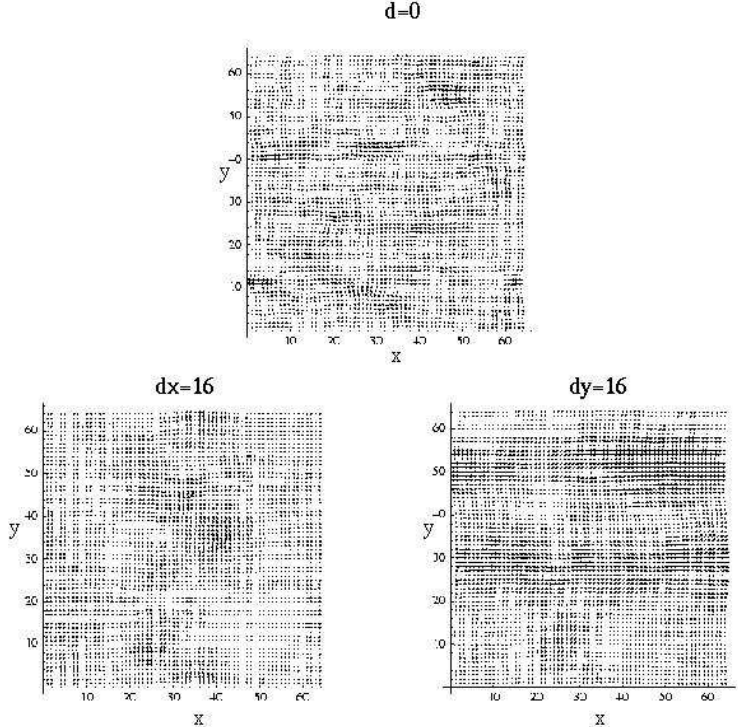}
\caption[shift_field]{\label{shift_field} The gap-averaged velocity field
$\delta u(x,y) = u(x,y)-U_x \check{x}$ is presented in three
different cases. On the top is the case with no lateral shift
between the surfaces of the fracture. On the bottom-left
we show the case $d=d_{\parallel}=16$, and the case
$d=d_{\perp}=16$ is shown on the bottom-right corner.
In all cases the vertical separation between surfaces is $h=16$.}
\end{figure}

In Fig.~\ref{shift_field} we show the effect of 
the orientation of the shift on the gap-averaged velocity
fluctuations, $\delta u(x,y) = u(x,y)-U_x \check{x}$, where the mean flow
is subtracted in order to magnify the fluctuations in the local velocity.
The three different cases presented in Fig.~\ref{shift_field}, 
$d=0$, $d=d_{\parallel}$ and $d=d_{\perp}$, have the same vertical gap $h$,
and correspond to the same fracture (same self-affine surface).
In the case where the lateral shift is perpendicular to the flow $d=d_{\perp}$, 
flow channels oriented in the direction of the imposed pressure drop, 
whereas these oriented channels are not present when the shift is along 
the flow direction $d=d_{\parallel}$. This effect is 
sometimes referred to as channeling \cite{auradou01}.


\section{Tracer dispersion}
\label{dispersion}

Several tracer-dispersion mechanisms are present in the transport of fluids
through porous media and the relative importance of these mechanisms strongly 
depends on the mean flow velocity \cite{dullien}. Let us briefly review here the
different dispersion mechanisms and their dependence on
the P\'eclet number Pe, which measures the ratio between convective
and diffusive effects and it is defined, for the fluid flow through fractures, by 
$\text{Pe}=h\,U_x/D_m$, where $D_m$ is the molecular diffusivity, $h$ is
the aperture of the fractures and $U_x$ is the mean flow velocity.
In the case of two-dimensional fractures there are only two 
contributions to tracer dispersion, molecular diffusion, which dominates at very 
low flow rates, $\text{Pe} \ll 1$, and is independent of Pe, $O(\text{Pe}^0)$, and 
Taylor dispersion, which is $O(\text{Pe}^2)$ and therefore becomes dominant at high 
flow rates $\text{Pe} \gg 1$ \cite{koplik93,gutfraind95,GK2}. On the other hand, 
in three-dimensional fractures another mechanism comes into play, that is the presence of 
spatial fluctuations in the velocity field. As discussed in the previous section, the effective 
aperture of the fracture is not constant, even in the case when the two 
complementary surfaces have no relative lateral shift. Hence,
it is clear that the fluctuations in the local effective aperture 
gives rise to velocity fluctuations, as it was shown in Sec.~\ref{fluctuations}. 
Moreover, in contrast with the two-dimensional case, these velocity fluctuations are present 
even after the local velocity is averaged over the gap of the fracture.
One of the main effects of the spatial fluctuations in the fluid velocity 
is that an initially-flat invasion-front of tracer particles 
will become increasingly distorted, resulting in its broadening in time, 
as it can be observed in Fig.~\ref{front}. This so induced 
geometric dispersion of tracer particles has been reported 
in previous studies of dispersion in fractures \cite{roux98,plouraboue98}
and is completely analogous to that observed in three-dimensional porous media 
\cite{saffman59}. However, let us note an important difference
between previous studies and the present work. In Ref. \cite{roux98,plouraboue98}
the analysis is based on the lubrication or Reynolds approximation,
where a Poiseuille flow, with a parabolic velocity profile across the aperture,
is assumed to be locally valid everywhere in the fracture. In this case,
there are no fluctuations in the local velocity in the absence
of fluctuations in the local aperture of the fractures, which is in fact the case when 
there is no lateral shift between surfaces.
On the other hand, we consider the case of narrow fractures and
the lubrication approximation is not valid (at least in its simplest version).
In fact, in our framework, geometric dispersion effects are present
even in the absence of any lateral shift between surfaces,
due to variations in the effective aperture that induce spatial
fluctuations in the velocity field.  This situation was investigated
in recent experiments, where the broadening and dispersive behavior of an invasion-front 
of tracer particles has been observed even in the case of no lateral shift between 
complementary surfaces \cite{auradou01}. 

In view of our previous discussion, we shall focus
on how the fluctuations in the gap-averaged velocity affect 
the dispersion of the tracer particles. 
The molecular and Taylor contributions to the dispersion 
of tracer particles were discussed in our previous work, in the
two-dimensional case, and they are not expected to be very sensitive to 
the fluctuations in the gap-averaged flow velocity, in that
the molecular diffusion is clearly independent of the velocity field
and the Taylor dispersion is dominated by the gradients in velocity in 
the direction perpendicular to the fracture surface \cite{GK2}. 
Then, if we only account for the geometric contribution to the dispersion
of tracer particles the problem can be immensely simplified. In fact, instead
of working with the three-dimensional velocity field we can use the two-dimensional
gap-averaged velocity field $u(x,y)$.
The range in which the geometric dispersion is the dominant mechanism
contributing to the dispersion of tracer particles corresponds to
intermediate P\'eclet numbers (intermediate velocities), and the presence of such
a range of P\'eclet numbers in self-affine fractures will be discussed
in detail at the end of this section (see Sec.~\ref{highPe}).


\subsection{Velocity autocorrelation function}
\label{autocorrelation}

\begin{figure}
\includegraphics[angle=-90,width=\W]{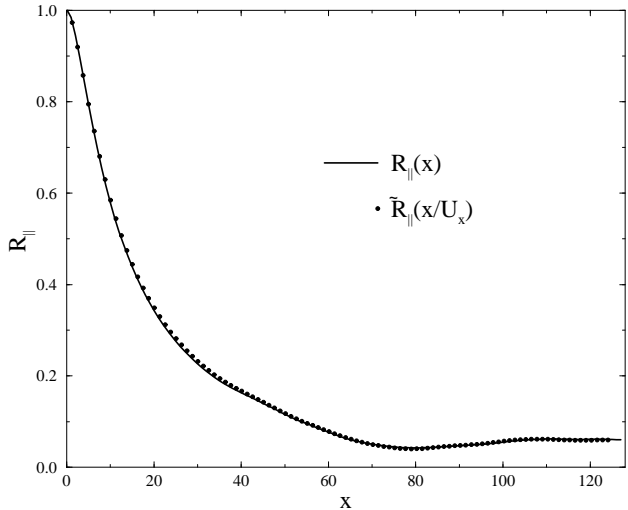}
\caption[correla]{\label{correlaTX} Spatial velocity-autocorrelation-function $R_{\parallel}(x)$ and
its comparison with $\tilde{R}_{\parallel}(x/U_x)$, both normalized by $\sigma_u^2$.
The results correspond to a system of size $L=256$ with the fracture
surfaces separated by $h=8$.}
\end{figure}

The dispersion coefficient in the direction of the flow can 
be expressed in terms of the velocity autocorrelation function 
in time $\tilde{R}_{\parallel}(t)$ \cite{reif},
\begin{eqnarray}
\label{DvsT}
\nonumber
D_H 
&=& \int_0^{\infty} \tilde{R}_{\parallel}(t) dt \\
&=& \lim_{T \to \infty} \int_{0}^{T} 
\left\langle [u_x(0)-U_x] [u_x(t)-U_x] \right\rangle dt,
\end{eqnarray}
where the integration is over the trajectory of the
tracer particles and the average is an {\it ensemble} average over 
different realizations of the problem. Furthermore, in Sec.~\ref{fluctuations} we showed 
that the velocity field is
approximately one dimensional, i.e. lateral
fluctuations are small compared to the mean velocity
($\delta u / U_x < 0.05$). 
Then, the time autocorrelation function of the velocity fluctuations,
$\tilde{R}_{\parallel}(t)$, can be related to the spatial 
velocity autocorrelation function, specifically to the marginal spatial
autocorrelation function in the direction of the mean flow $x$ 
integrated over the $y$ direction,
\begin{equation}
R_{\parallel}(x) = \left\langle [u_x(x',y')-U_x] [u_x(x'+x,y')-U_x] \right\rangle,
\end{equation}
by transforming time into position through the mean flow 
velocity, $x=U_x~t$. In Fig.~\ref{correlaTX} we show the 
equivalence between $R_{\parallel}$ and $\tilde{R}_{\parallel}$, that is
$R_{\parallel}(x) = \tilde{R}_{\parallel}(x/U)$. Then, we can rewrite the dispersion coefficient 
in terms of the spatial correlation function,
\begin{equation}
\label{DvsX}
D_{H} = \frac{1}{U}\int_{0}^{\infty} R_{\parallel}(x) dx.
\end{equation}

From the previous equation it is clear that the dispersion coefficient depends
on both the magnitude of the fluctuations and the length scale of the velocity correlations. 
Therefore, we introduce the characteristic correlation length of the velocity 
fluctuations, $l_c$, defined as
\begin{equation}
\label{correlaX}
l_{c} = \frac{1}{\sigma^2_u} \int_{0}^{\infty} R_{\parallel}(x) dx, 
\end{equation}
where $\sigma_u$ is the rms dispersion in velocity, 
$\sigma^2_u = \left\langle (u_x-U_x)^2 \right\rangle$. 
The velocity correlation length $l_{c}$ measures
the typical length over which fluctuations in velocity 
are correlated and, similarly, we can define a correlation time 
$\tau_c=l_c/U_x$, which measures the characteristic time scale 
over which fluctuations in velocity remain correlated. Let us remark that $l_{c}$
is not necessarily equal to the previously defined linear size of the 
quasi-linear blocks $\xi$, a fact that becomes clear upon consideration
of the 2d case, where one can define a typical size $\xi$ over which the channel
formed by the opposing fracture surfaces can be considered straight,
even though there are no fluctuations in the mean velocity and,
therefore, the correlation length $l_c$ cannot be defined. 

In an analogous way, we might define the velocity correlation length 
for the spatial correlations in the direction perpendicular to the mean flow,
\begin{eqnarray}
\label{correlaY}
\nonumber
l_{\perp} &=& \frac{1}{\sigma^2_u} \int_{0}^{\infty} 
\left\langle (u_x(x,0)-U_x)(u_x(x,y)-U_x) \right\rangle dy \\ 
&=& \int_{0}^{\infty} R_{\perp}(y) dy. 
\end{eqnarray}
However, it can be shown that this integral is exactly zero, due 
to mass conservation, which suggests the presence of negative spatial correlations 
in the direction perpendicular to the flow.

\begin{figure}
\includegraphics[angle=-90,width=\W]{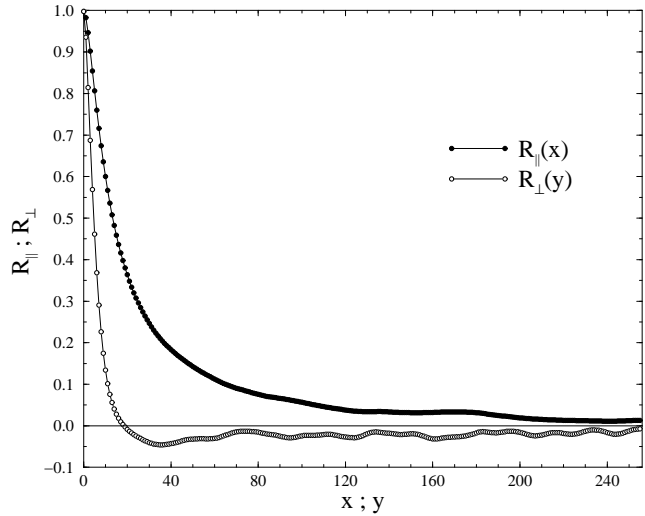}
\caption[correlaxy]{\label{correlaXY} Spatial velocity-autocorrelation-function, 
of the velocity component in the direction of the flow $u_x-U_x$,
in both the direction of the flow $R_{\parallel}$ and perpendicular 
to it $R_{\perp}$. The autocorrelation functions are normalized by $\sigma_u^2$.
The results correspond to a system of size $L=512$ and fracture
surfaces separated by $h=8$.}
\end{figure}

In Fig.~\ref{correlaXY} we present both $R_{\parallel}$ and 
$R_{\perp}$ for a system of size $L=512$ and gap size $h=8$. 
It is clear that the spatial correlations in the direction 
perpendicular to the mean flow decay faster than along the flow direction, which is 
consistent with our previous result concerning the magnitude of the fluctuations, 
in that velocity fluctuations parallel to the mean flow are larger than 
fluctuations perpendicular to it (see Sec.\ref{fluctuations}). 
It can also be observed that both correlation functions
do not vanish at long distances as it should be in an
infinite system. In fact, the velocity fluctuations present a positive 
correlation in the direction of the flow and are anti-correlated 
in the perpendicular direction. 
This non-vanishing correlation can be explained in terms of
mass conservation and finite size effects.
The gap-averaged velocity $u_x$ integrated over a line perpendicular to
the mean flow (from $y=0$ to $y=L$) is equal to the
total flow rate $Q$ divided by the gap size $h$, and it 
is a conserved quantity all along the system. 
Therefore, local fluctuations in the velocity $u_x$ should
compensate themselves, giving rise to negative spatial correlations 
in $u_x$ along the direction perpendicular to the flow. 
However, this asymptotic negative correlation should decrease as the
size of the system increases, as it can be observed in our results
by comparing Fig.~\ref{correlaTX}, which corresponds to a system of size $L=64$,
with Fig.~\ref{correlaXY}, which corresponds to $L=256$. 
This same combination of effects, that is mass conservation and 
the finite size of the system, leads to the positive correlation
in the direction of the flow observed in Fig.~\ref{correlaXY}. 
The negative spatial correlation in the velocity field along the direction perpendicular
to the mean flow, and the fact that the
velocity fluctuations decay faster in this direction, are
also found in the continuum approach to transport in
heterogeneous porous media \cite{dagan89}.

\subsection{Dependence of dispersion on the gap size}
\label{D_vs_h}

Let us show first that, as it was assumed in the introduction 
to Sec.~\ref{dispersion}, the contribution to the dispersion coefficient
due to the spatial fluctuations in the velocity is,
for vanishingly small Reynolds numbers Re, linear in the mean flow 
velocity $U_x$. In the Stokes flow approximation ($\rm{Re}=0$), the flow field
becomes independent of the magnitude of the flow rate, $U_x$
being just a scaling factor \cite{batchelor}. Therefore,
the dimensionless parameter $\delta = \sigma_u/U_x$, and the
correlation length $l_c$, are independent of $U_x$. Then, 
rewriting Eq.~(\ref{DvsX}) we obtain,
\begin{equation}
\label{D}
D_H = \delta^2~l_{c}~U_x,
\end{equation}
which explicitly shows the linear dependence of the dispersion
coefficient on $U_x$. Furthermore,
the dispersion of an initially-flat front of tracer particles 
at a given distance from the injection point
is independent of $U_x$. Consider the situation in which a front of tracer particles 
is injected at the inlet section of a fracture ($x=0$).
The dispersion of the tracer front, measured as the mean square displacement 
of the tracer particles, is then given by,
\begin{equation}
\label{dx}
\left\langle (x - \left\langle x \right\rangle)^2 
\right\rangle_t = 2 \, D_H \, t = 2 \, \delta^2 \, l_c \, U_x \, t =
2 \, \delta^2 \, l_{c} \, \left\langle x \right\rangle_t,
\end{equation}
and it is clear that the dispersion of the front, at a fixed distance 
$x=\langle x \rangle$ from the inlet section $x=0$, is independent of $U_x$.
In fact, it only depends on $l_D=\delta^2 \, l_c$, where $l_D$ is
the dispersion length of the fracture \cite{bacri87}. 
This fact, that the dispersion of the front is independent of the mean flow velocity,
was observed in the experiments presented in Ref. \cite{auradou01}, 
where it is shown that the front shape depends on the injected volume but 
not on the flow rate, (see Fig.~2 in \cite{auradou01}).

Let us then, investigate the dependence of the dispersion length $l_D$
on the gap size $h$, accounting the dependence on $h$ of both the 
relative magnitude of the fluctuations $\delta$ and the correlation 
length $l_c$. 

The dependence of $\delta$ on the gape size was 
discussed in Sec.~\ref{fluctuations}, where we showed that,
\begin{equation}
\label{delta}
\delta^2 \propto h^{-4(1-\zeta)/\zeta}.
\end{equation}

\begin{figure}
\includegraphics[angle=-90,width=\W]{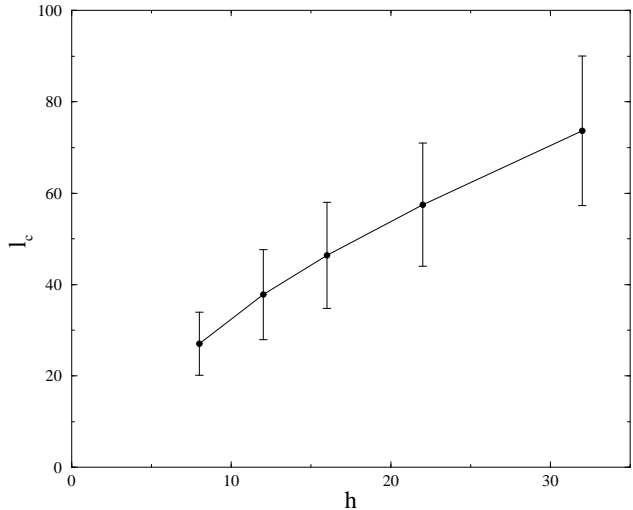}
\caption[length]{\label{correla_length} Correlation length $l_c$ as a function 
of the gap size $h$. The size of the system is $L=256$.}
\end{figure}

On the other hand, in Fig.~\ref{correla_length} we show the correlation 
$l_c$ as a function of the gap size, computed from our numerical 
simulations using Eq.~(\ref{correlaX}), where it can be seen that 
the correlation  length increases with the gap size $h$.
Let us mention that, in order to minimize finite-size effects, as 
the previously discussed non-vanishing spatial correlations in the velocity 
fluctuations, present in both parallel and transverse directions,
the computation of $l_c$ was performed in the largest system
we could simulate, that is $L=1024$.

The observed decrease in the spatial correlations of the velocity field, as 
the surfaces become closer, might be attributed to a `screening'
mechanism, that is, as $h$ decreases the fluctuations in the velocity field
become stronger and the velocity tends to decorrelate over a shorter distance.
An analogous effect occur in porous media flows,
where a velocity disturbance from a point force 
decays with a characteristic `screening' length
proportional to the square root of the permeability
(as seen from the Brinkman equation \cite{brinkman47,durlofsky87}),
which in our case is proportional to $h$ 
(the leading order term).

\begin{figure}
\includegraphics[angle=-90,width=\W]{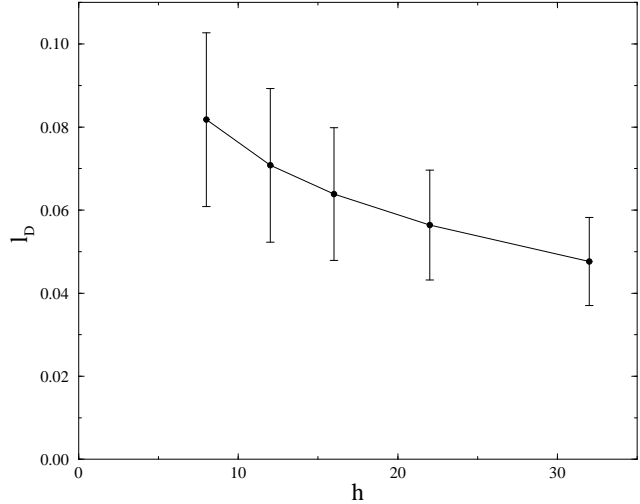}
\caption[dispersion]{\label{Dx_vs_h} Dispersion length $l_D$ as a function of the gap size $h$. 
The size of the simulated fractures is $L=1024$.}
\end{figure}

As discussed earlier, the dependence of $l_D$ on $h$ has two opposite
contributions. On one hand, the magnitude of the velocity fluctuations,
$\delta$, which decreases with $h$ and, on the other hand, the correlation length
of the velocity fluctuations, $l_c$, which becomes larger as $h$ increases.
We found that, as a result of this
opposite effects, the dispersion length decreases as
the gap size is increased, as it can be seen in 
Fig.~\ref{Dx_vs_h}, where we show the dispersion length $l_D$
as a function of the gap size $h$. This result is in agreement
with the qualitative behavior observed in Ref.~\cite{auradou01},
where the invasion front of tracer particles becomes smoother
as the gap of the fracture is increased (see Fig.~5 in \cite{auradou01}.)

\begin{figure}
\includegraphics[angle=-90,width=\W]{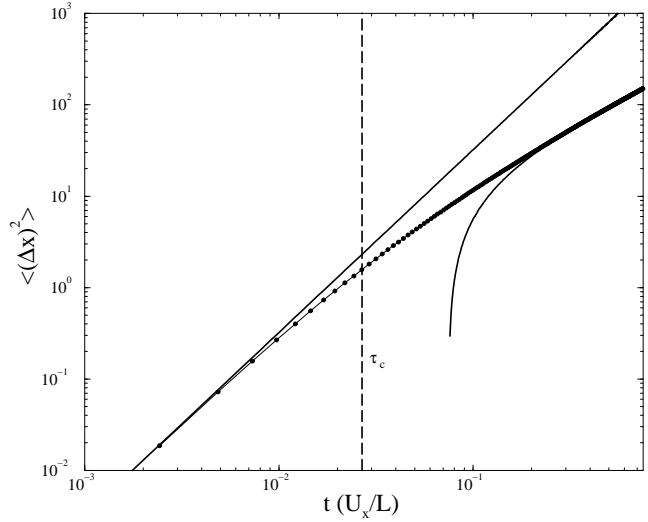}
\caption[xx]{\label{xx} Log-log plot of the mean square displacement of an initially flat 
front of tracer of particles as a function of time. The vertical dashed line corresponds 
to the dimensionless correlation time $\tau_c=l_c/L$ above which a diffusive behavior is
expected. The upper solid line corresponds to the initial highly-correlated
motion of the tracer particles, $\left\langle (\Delta x)^2\right\rangle=\sigma_u^2 t^2$.
The lower solid line corresponds to the best fit
of the linear regime, with a dispersion coefficient $D_H=(2.7 \pm 0.7)\times10^{-5}$.
(The observed departure from a linear behavior at early times is due to the 
constant term of the fitted linear regime, $(\Delta x)^2_{t=0} \sim -15$. )
The results correspond to
simulation in a system with $L=1024$ and gap size $h=8$, and
were averaged over 4 different realizations. The time is 
in units of the mean transit time of the medium $T=L/U_x$.}
\end{figure}

Finally, note that underlying the previous discussion
is the assumption that the spatial correlations in the
velocity field decay fast enough so that the integral 
in Eq.~(\ref{DvsT}) is finite and, therefore, the broadening of the 
tracer front becomes diffusive at length scales larger than the 
correlation length $l_c$. Analogously, the dispersion of the front 
is expected to be diffusive at time scales larger than the correlation 
time $\tau_c$. On the other hand, in previous studies of dispersion
in self-affine fractures, the slow decay in the spatial correlations of the 
velocity field was proven to induce anomalous dispersion. However,
they investigated the dispersion of tracer in the lubrication regime, where
mean velocity strictly follows the fluctuations in the aperture of the
fracture and long-range correlations should be expected. On the contrary, the
lubrication approximation is not valid in the case of {\it narrow fractures},
and fluctuations in the velocity field are due mainly to the locally random
orientation of the fracture channel. In Fig.~\ref{xx}
we present the mean square displacement of an invasion front of
tracer particles as a function of time. The vertical line shows the
correlation time, $\tau_c$, after which a diffusive behavior should
be expected. It can be observed that the initial behavior corresponds
to the highly correlated motion of the particles and, in fact, the numerical 
results closely follows the solid line that is given by 
$\left\langle (\Delta x)^2\right\rangle=\sigma_u^2 t^2$.
On the other hand, at times larger than $\tau_c$ the velocity begins to
decorrelate from its previous values, and the dispersion of the front
deviates from the quadratic behavior. Moreover, the lower solid line
is given by a linear fit to the mean square displacement, in the range
of times $0.2 \leq t (Ux/L) \leq 0.75$. The fitted value for the dispersion coefficient is
$D_H=(2.7 \pm 0.7)\times10^{-5}$, in agreement with the expected value
calculated from Eq.~(\ref{D}), $D_H=(2.0 \pm 0.6)\times10^{-5}$ 
However, the size of the system is not large enough to observe a large range
where the linear, dispersive, regime is valid, and therefore
the determination of the dispersion coefficient is not accurate.
Nevertheless, the results allow us to conclude that the 
mean square displacement grows at a much slower rate than in the anomalous regime
observed in Refs.~\cite{roux98,plouraboue98}, where $(\Delta x)^2\propto t^{2\zeta}$.

\subsection{Tracer dispersion in the three-dimensional velocity field and
tracer transit time distributions}
\label{highPe}

The previous analysis of the dispersion problem 
was based on the assumption that the leading 
contribution to the dispersion of tracer particles comes 
from the spatial fluctuations in the gap-averaged velocity field, which allowed
us to map the problem to a two-dimensional one. 
This approximation is only valid for values of the P\'eclet numbers
such that both molecular and Taylor dispersion are negligible 
\cite{plouraboue98}. As we shall discuss, there might be not such a
range of P\'eclet numbers, depending on the geometric
properties of the fracture.

The geometric contribution to the dispersion coefficient,
given by Eq.~(\ref{D}), is larger than the molecular 
diffusion term whenever the following inequality holds,
\begin{equation}
\label{DG.lt.DM}
D_H = \delta^2 l_c U_x \gg D_m \Longrightarrow \text{Pe} \gg \frac{1}{\beta} 
\qquad \beta=\delta^2 \frac{l_c}{h}.
\end{equation}

On the other hand, we might expect that Taylor-like
dispersion becomes dominant at high flow rates, due to the presence
of stagnant zones whithin the fracture. In that case, a heuristic 
estimate of the range of Peclet numbers where
the geometric contribution generates a larger spreading
of the tracer front than that induced by the presence of stagnant zones 
is given by,
\begin{equation}
D_H = \delta^2 l_c U_x \gg D_T=\frac{h^2U_x^2}{D_m} \Longrightarrow
\text{Pe} \ll \beta,
\end{equation}
Then, combining the previous to equations, it is clear that
the geometric regime exists only for
\begin{equation}
\beta = \delta^2 \frac{l_c}{h} \gg 1
\end{equation}
That is, the product between the magnitude of the fluctuations
in velocity and the characteristic length over which such fluctuations
remain correlated should be large. 
Therefore, the existence of such a range of P\'eclet numbers
in which the dispersion of tracer particles due to the velocity fluctuations
is dominant, would depend on the geometrical properties of the fracture.
In view of our previous results, in particular the dependence of $\delta$ and
$l_c$ on $h$, we might expect that the geometric dispersion
would be dominant in the limit of {\it narrow fractures}, i.e. large fluctuations
of the surface height and small separation between fracture
surfaces. In terms of the small parameter $\epsilon=\sigma_z(\xi)/\xi$,
the geometric contribution will be asymptotically dominant, for
any value of the P\'eclet number, in the limit $\epsilon \to 0$. On the other hand,
as the gap of the fracture is increased, the geometric
contribution will be asymptotically negligible in the
limit $h \to \infty$.

Finally, let us consider the transit time distribution of tracer particles
at high injection rates. Previously, we have analyzed the fully-developed
dispersion regime, which is valid at low injection rates or large fractures. 
Specifically, let us measure the transit time of tracer 
particles, launched at the inlet section of the fracture, that
arrive at the cross-section situated at a distance $\Delta X$.
If $T$ is the mean transit time of the tracer particles, $T=\Delta X/U_x$,
and $\tau_D$ is the characteristic correlation time 
of the tracer velocity, then the Gaussian dispersive behavior
would be valid for $T\gg\tau_D$.
On the other hand, when the injection rate increases and the transit time becomes 
$T\lesssim\tau_D$, the transit time distribution deviates from a Gaussian distributions 
and exponential tails are generally observed in flow through porous media 
\cite{coats,plouraboue98,chem2}. 

\begin{figure}
\includegraphics[angle=-90,width=\W]{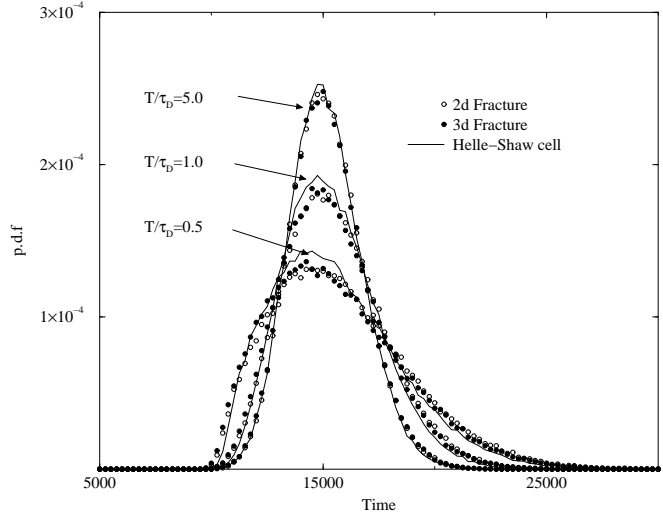}
\caption[dispersion]{\label{disp3D} Transit time distribution
of tracer particles for three different
ratios between the transit and correlation times,
$T/\tau_D = 1/2, 1~\text{and} 5.0$, and for three different systems,
a Hele-Shaw cell, 2d and 3d fractures. The simulation correspond
to a system of size $L=512$, gap size $h=16$ and a transit time
distributions are measured at a distance $\Delta X= 400$ from the injection
point.}
\end{figure}

In the flow through fractures, as well as in the flow 
in a Hele-Shaw cell, the correlation 
time of the velocity is given by the diffusive time across the gap. For transit times 
$T\lesssim\tau_D$ the tracer particles do not have time to diffuse 
across the gap and their velocity will remain correlated during their 
convective motion throughout the system. In this case, stagnant or 
low-velocity zones have the effect of retarding the tracer particles,
and give rise to the exponential tails, due to the fact that diffusive 
motion is the only mechanism available for the particles to leave 
these stagnant zones \cite{bouchaud}. In Fig.~\ref{disp3D}
we present the tracer transit time distribution for three different
ratios between the transit and the correlation times,
$T/\tau_D = 1/2, 1~\text{and}~5.0$, and for three different systems,
Hele-Shaw cells, 2d and 3d fractures. First of all, it can be 
observed that, for $T/\tau_D \gg 1$, all the transit time 
distributions are Gaussian and very similar to each other.
On the other hand, as the transit times becomes  $T/\tau_D \lesssim 1$,
the distributions deviate from a Gaussian curve, becoming
increasingly asymmetric. 
It can also be observed that both two and three dimensional fractures
present slightly more persistent tails than in the Hele-Shaw case.
The similarity between the
distribution in $2d$ and $3d$ fractures is in total agreement with our previous results,
where the velocity field for three-dimensional fractures was shown
to be quasi-two-dimensional in the absence of a lateral shift between the surfaces of the
fracture. On the other hand, we feel that the small difference between
the transit time distribution in self-affine fractures
and in the Hele-Shaw cell might be related to the presence of
low-velocity zones in the fractures, which are not present in a straight
channel. This enhancement of the long-time tails
due to the presence of low-velocity zones should be more
evident in the presence of a lateral shift.
\begin{figure}
\includegraphics[angle=-90,width=\W]{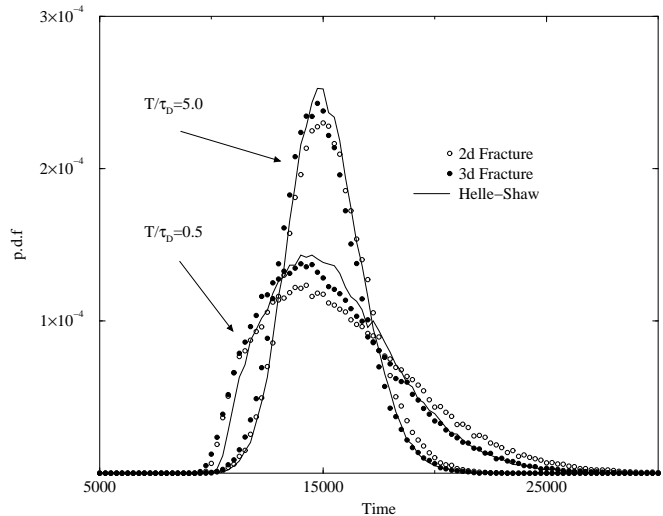}
\caption[dispersion]{\label{disp3D_dx} Transit time distribution
of tracer particles for two different
ratios between the transit and correlation times,
$T/\tau_D = 1/2 ~\text{and} 5.0$, and for three different systems,
a Hele-Shaw cell, 2d and 3d fractures with opposite surfaces
laterally displaced by $d=dx=16$. The simulation correspond
to a system of size $L=512$, gap size $h=16$ and a transit time
distributions are measured at a distance $\Delta X= 400$ from the injection
point.}
\end{figure}
In Fig.~\ref{disp3D_dx} we present the transit time distribution
of tracer particles, in the same three different cases as in Fig.~\ref{disp3D},
but in this case the upper surface of the fractures is laterally 
shifted in the direction of the flow by $d=dx=16$. 
First of all, it can be observed that, as in the unshifted case presented in Fig.~\ref{disp3D},
all distributions are Gaussian and very similar to each other for large transit times,
$T/\tau_D = 5.0$.
On the other hand, for much smaller transit times, $T/\tau_D = 0.2$,
a long-time tail develops, in particularly in the case of two-dimensional fractures.
As shown in Sec.~\ref{shift}, two-dimensional fractures present lower
permeabilities than the three-dimensional ones in the presence of a lateral shift,
due to the fact that in the $3d$ case the fluid can avoid low
permeability regions by flowing around them. Therefore,
the presence of low-velocity zones is more important in $2d$ 
and thereby the effect of these zones on the long-time behavior of the transit 
time distribution becomes more important.

\section{Summary and conclusions}
\label{summary}

Transport properties of three-dimensional self-affine rough 
fractures were investigated by means of the effective-medium
approximation and numerical simulations using the Lattice-Boltzmann
method. The numerical simulations verified the scaling behavior predicted
by the effective-medium approach and, furthermore, allowed us to compute
important transport parameters of the fractures, such as the dispersion length,
and their dependence on the size of the aperture. Two different cases
were investigated, the unshifted case, in which the two
matching surfaces of the fracture are displaced in the direction
normal to the mean plane, and the shifted one,
in which the upper surface is laterally displaced either in the direction
of the flow or perpendicular to it.

First, we modeled the fractures by a regular two-dimensional square-lattice
of bond conductances, and the lattice spacing and the distribution of 
bond conductivities were related to the geometrical properties of the fracture. Specifically,
we related the lattice-spacing to the length scale over which fluctuations
in the surface height are small compared to the aperture of the fracture,
and it was determined in terms of the roughness exponent, the characteristic length 
and the aperture of the fractures. Then, adapting some well-known
results obtained by means of the effective-medium approximation 
in the analogous random-resistor network, we obtained the permeability
of the fracture and its dependence on the aperture size in the limit
of {\it narrow fractures}. We showed that the permeability is,
up to second order in a perturbative parameter, the same as in 
two-dimensional fractures. This quasi two-dimensional behavior of the 
transport of fluids through self-affine fractures was confirmed by the numerical 
computations of the streamlines, which presented very small lateral fluctuations. 
A similar behavior was also observed in the experimental work reported in 
Ref.~\cite{auradou01}, in that the structure developed by the invasion front
of tracer particles presents very small fluctuations perpendicular to the
mean flow in the unshifted case (see Fig.~4a in Ref.~\cite{auradou01}).
Moreover, the scaling behavior of the permeability 
with the aperture size was verified by our numerical
results and, in addition to that, we showed that it is in agreement with numerical 
results performed in two-dimensional fractures. However, we also discussed an important
difference between the 3d and 2d with cases, namely, the presence of 
fluctuations in the gap-averaged fluid velocity in three-dimensional fractures, 
and we were further able to predict the scaling behavior of the velocity fluctuations 
in the direction
of the mean flow by means of the effective-medium approximation. The numerical
simulations were in agreement with this result and, furthermore, showed that the
velocity fluctuations in the direction perpendicular to the
flow have the same scaling but are approximately three times smaller in magnitude. 
Similar results were obtained in the continuum approach to
transport in heterogeneous porous media \cite{dagan84,dagan89}. Finally, we investigated
the case of shifted surfaces and showed that the permeability of the fractures
strongly depends on the orientation of the shift, which is either in the direction 
of the imposed pressure drop or perpendicular to it, in the limit
of {\it narrow fractures}. Furthermore, by means of numerical simulations we showed 
that the flow rate per unit width in three-dimensional fractures is bounded by the 
two-dimensional results. Specifically, for a relative shift in the direction
perpendicular to the mean flow the permeability is slightly affected and
lies above the permeability of two-dimensional fractures. On the other hand,
when the upper surface is shifted in the direction of the flow the permeability
is largely reduced, but not as much as in the two-dimensional case. The latter
is due to the fact that in three-dimensional fractures, 
in contrast with the two-dimensional case, the fluid 
can avoid low-permeability regions by flowing around them. We also presented
a simplified representation of a local region of the fracture that, although naive in
character, captures the effect of the orientation of the shift on the
permeability of the fracture. 
The same effect is observed in experiments, in that the flow rate is larger in the direction
perpendicular to the relative shift between surfaces 
(see Fig.~4 in \cite{auradou01}).

In the second part of this work, we investigated the dispersion of tracer particles
in self-affine fractures. Specifically, we analyzed the dependence of the
geometric contribution to the dispersion process on the aperture size.
First, we simplified the analysis by mapping the problem to the dispersion
of tracer particles in the two-dimensional gap-averaged velocity field. 
We then distinguised between the two contributions to the dispersion coefficient,
namely the relative magnitude of the velocity fluctuations and their
correlation length. We observed that, in agreement with previous studies \cite{dagan84,dagan89},
the autocorrelation function of the velocity fluctuations
decays faster in the direction perpendicular to the mean flow. 
Finally, we showed that,
even though the correlation length increases with the size of the aperture,
the dispersion coefficient is asymptotically small in the limit of wide fractures.
The latter effect is also observed in the experiments presented in 
Ref.~\cite{auradou01} were it was shown that the invasion front of tracer
particles becomes increasingly smooth as the aperture of the fracture is 
increased.

Finally, we investigated the dispersion of tracer particles
in the fully three-dimensional velocity field inside the fractures.
Specifically, we discussed the range of P\'eclet numbers in which
the geometric contribution to dispersion is expected to be dominant and we
showed that, depending on the geometric properties of the
fracture, there might be no such a range of P\'eclet numbers.
However, we found that the geometric dispersion is dominant
in the limit of {\it narrow fractures}, and in general
is dominant for large relative fluctuations of the velocity field
and correlation lengths larger than the aperture of the fractures.
We also investigated the transit time distribution of tracer particles,
and their dependence on the mean transit time. We showed that, as the
mean transit time is reduced and it becomes comparable
or smaller than the correlation time of the tracer velocity,
the transit time distribution becomes increasingly non-Gaussian,
developing long-time tails due to the presence of low-velocity
zones where the only mechanism for tracer transport is molecular diffusion.
In general, the transit time distributions
are very similar to the case of tracer dispersion in a Hele-Shaw cell,
except for the case of two-dimensional fractures shifted in the direction
of the flow, which present the largest tails probably due to
an enhancement of the low-velocity zones by the relative shift between surfaces. 
In fact, in agreement with the latter results, the two-dimensional 
fractures, with the upper surface shifted in the direction of the flow,
were shown to have the lowest permeability.

\section*{Acknowledgments}

We thank J. P. Hulin and H. Auradou for useful discussions
and for sharing with us their experimental results; 
G. Dagan for carefully reading the manuscript and for his many
suggestions; R. Chertcoff, I. Ippolito, D. L. Johnson and N. Nerone
for useful discussions. 
This research was supported by the Geosciences Research Program, Office of
Basic Energy Sciences, U.S. Department of Energy, and computational facilities
were provided by the National Energy Resources Scientific
Computer Center.

\end{document}